\documentclass[aps,prd,final,twocolumn,letterpaper,superscriptaddress,footinbib,notitlepage]{revtex4-1}

\usepackage{graphicx}                       
\usepackage[outdir=./]{epstopdf}
\usepackage{amsmath}
\usepackage{amssymb}
\usepackage{amsthm}
\usepackage{dsfont}       
\usepackage[usenames,dvipsnames,table]{xcolor}
\usepackage[format=plain, font=small, labelfont=bf, justification = raggedright]{caption}
\usepackage{array}  
\usepackage{overpic}
\usepackage{placeins} 
\usepackage{fancyhdr}
\usepackage{colortbl}
\usepackage{multirow} 
\usepackage{hyperref} 
\hypersetup{colorlinks = true,linkcolor = blue, breaklinks = true}

\newcommand{\eq}[1]{(\ref{#1})}
\newcommand{\Eq}[1]{Eq.~\eq{#1}}
\newcommand{\Eqs}[1]{Eqs.~\eq{#1}}
\newcommand{\Fig}[1]{Fig.~\ref{#1}}
\newcommand{\Sec}[1]{Sec.~\ref{#1}}

\renewcommand{\Ref}[1]{Ref.~\cite{#1}}

\newcommand{\Refs}[1]{Refs.~\cite{#1}}
\newcommand{\App}[1]{Appendix~\ref{#1}}

\newcommand{\eg}{{e.g., }}
\newcommand{\ie}{{i.e., }}

\newcommand{\mc}[1]{\mathcal{#1}}

\newcommand{\msf}[1]{\mathsf{#1}}

\newcommand{\mbb}[1]{\mathbb{#1}}

\newcommand{\bra}[1]{\langle#1 |}
\newcommand{\ket}[1]{|#1 \rangle}

\newcommand{\fourier}[1]{\smash{\widetilde{#1}}}

\newcommand{\pd}[1]{\partial_{#1}}

\newcommand{\dd}{\mathrm{d}}

\DeclareMathOperator{\airyA}{Ai}
\DeclareMathOperator{\airyB}{Bi}

\newcommand{\Vect}[1]{{\boldsymbol{\rm #1}}}

\newcommand{\Mat}[1]{\msf{#1}}

\newcommand{\Symb}[1]{\mc{#1}}

\newcommand{\cont}[1]{\mc{C}_{#1}}

\newcommand{\MTnorm}{\mc{N}_\Vect{t}}

\newcommand{\nullFrac}{\vphantom{\frac{}{}}}

\newcommand{\Stroke}[1]{\text{\ooalign{ $#1$\cr \hidewidth\raise.225ex \hbox{$-\mkern.5mu$}\cr}}}

\begin{document}
\setlength{\parskip}{0pt}
\setlength{\belowcaptionskip}{0pt}


\title{Steepest-descent algorithm for simulating plasma-wave caustics via metaplectic geometrical optics}
\author{Sean M. Donnelly}
\affiliation{Department of Physics \& Astronomy, Iowa State University, Ames, Iowa 50011, USA}
\author{Nicolas A. Lopez}
\affiliation{Department of Astrophysical Sciences, Princeton University, Princeton, New Jersey 08544, USA}
\author{I.~Y. Dodin }
\affiliation{Department of Astrophysical Sciences, Princeton University, Princeton, New Jersey 08544, USA}
\affiliation{Princeton Plasma Physics Laboratory, Princeton, New Jersey 08543, USA}

\begin{abstract}
The design and optimization of radiofrequency-wave systems for fusion applications is often performed using ray-tracing codes, which rely on the geometrical-optics (GO) approximation. However, GO fails at wave cutoffs and caustics. To accurately model the wave behavior in these regions, more advanced and computationally expensive ``full-wave" simulations are typically used, but this is not strictly necessary. A new generalized formulation called metaplectic geometrical optics (MGO) has been proposed that reinstates GO near caustics. The MGO framework yields an integral representation of the wavefield that must be evaluated numerically in general. We present an algorithm for computing these integrals using Gauss--Freud quadrature along the steepest-descent contours. Benchmarking is performed on the standard Airy problem, for which the exact solution is known analytically. The numerical MGO solution provided by the new algorithm agrees remarkably well with the exact solution and significantly improves upon previously derived analytical approximations of the MGO integral.
\end{abstract}

\maketitle

\pagestyle{fancy}
\lhead{Donnelly, Lopez, \& Dodin}
\rhead{Numerical MGO}
\thispagestyle{empty}

\section{Introduction}

Electromagnetic (EM) waves are widely used in plasma applications, including magnetic confinement fusion~\cite{Stix92,Freidberg10,Wesson11} and inertial confinement fusion~\cite{Lindl04,Craxton15}. Accurately modeling how EM waves propagate in plasma is therefore of upmost importance. Full-wave modeling, that is, directly solving Maxwell's equations with appropriate source terms, can be computationally expensive. Instead, the geometrical-optics (GO) approximation is often used to quickly calculate the wave amplitude along the GO rays that illuminate the region of interest~\cite{Tracy14, Kravtsov90}. The obtained amplitude profile can then be used as a source term in calculations of macroscopic plasma equilibrium~\cite{Prater08,Poli18a}. Design studies for EM-wave systems are often performed in this manner~\cite{Poli18a,Poli15,Poli16,Lopez18a}.

Unfortunately, GO solutions develop singularities at caustics such as cutoffs or focal points~\cite{Kravtsov93,Berry80b}. This is an issue for applications in which caustics play a central role, such as initializing spherical tokamak plasmas~\cite{Peng86,Peng00,Ono15a} via electron cyclotron resonance heating~\cite{Erckmann94,Prater04} (where the time-evolution of caustic surfaces directly defines the window of operation~\cite{Lopez18a}), or driving current in overdense plasmas via mode conversion to the electron Bernstein wave~\cite{Ram00,Shiraiwa06,Uchijima15,Seltzman17,Lopez18b,Laqua07,Preinhaelter73,Hansen85,Mjolhus84,Laqua03,Shevchenko07} (where the field structure near the EM wave cutoffs must be precisely resolved to obtain accurate mode-conversion efficiency estimates). Reduced modeling of these processes requires a more advanced machinery than traditional GO.

In response to this need, a new reduced theory called metaplectic GO (MGO) has been recently developed that leads to solutions that are finite at caustics~\cite{Lopez20a,Lopez21a}. By default, MGO yields an integral representation of the wavefield, which can be approximated analytically to some extent but in general must be evaluated numerically. Unfortunately, the integrands in MGO are highly oscillatory, so standard integration methods are insufficient~\cite{Deano17}. Special numerical algorithms tailored to MGO are needed. 

As part of ongoing work on MGO algorithms~\cite{Lopez20a,Lopez21a,Lopez19a,Lopez21b}, here we present a quadrature rule for calculating MGO integrals based on numerical steepest-descent integration~\cite{Deano09}. This algorithm emerges naturally from the MGO framework in that MGO integrals always contain saddlepoints that correspond to the ray contributions to the wavefield. We benchmark our algorithm on a class of examples in which the MGO integral contains a single isolated saddlepoint of various degeneracy, physically representing a wavefield either far from a caustic or at the critical point of a cuspoid-type caustic. We then apply our algorithm to the classic problem of an EM-wave reflecting off an isolated cutoff as governed by Airy's equation~\cite{Stix92,Tracy14}. We show that the numerical MGO solution agrees with the exact result amazingly well, much better than the analytical approximation to the MGO integral that was previously derived~\cite{Lopez20a}.

This paper is organized as follows. In \Sec{sec:background} the basic machinery of MGO is summarized, and the various types of caustics one expects to encounter are briefly surveyed. Section~\ref{sec:algorithm} constitutes the bulk of our paper, first introducing steepest-descent integration and Gaussian quadrature, then proceeding to derive our new quadrature rule. Benchmarking examples are provided in \Sec{sec:examples}, and \Sec{sec:concl} summarizes our main results.


\section{Metaplectic geometrical optics and caustics}
\label{sec:background}


\subsection{A brief overview}

Here we provide a brief overview of MGO; for more details, see \Refs{Lopez20a,Lopez21a}. Let $\psi(\Vect{x})$ be a scalar stationary wavefield in a plasma described by an $N$-dimensional ($N$-D) Euclidean coordinate system. (Generalizations to arbitrary metric are discussed in \Ref{Dodin19}). Neglecting nonlinear effects, the governing wave equation for $\psi(\Vect{x})$ is most generally written in the following integral form:
\begin{equation}
    \int \dd \Vect{x}' \,
    D(\Vect{x}, \Vect{x}')
    \psi(\Vect{x}')
    = 0
    ,
\end{equation}

\noindent where the integral kernel $D(\Vect{x}, \Vect{x}')$ is determined from the linear dielectric response of the plasma $\epsilon(\Vect{x}, \Vect{x}')$ in a known manner~\cite{Tracy14,Stix92}. For example, transverse EM waves have $D(\Vect{x}, \Vect{x}')$ given by
\begin{equation}
    D(\Vect{x}, \Vect{x}')
    = 
    - \nabla^2 \delta(\Vect{x} - \Vect{x}')
    - \epsilon(\Vect{x}, \Vect{x}')
    ,
\end{equation}

\noindent where $\nabla^2$ is the Laplacian operator with respect to $\Vect{x}$.

In the traditional GO limit, when $\psi(\Vect{x})$ is highly oscillatory, adopting the eikonal partition
\begin{equation}
    \psi(\Vect{x})
    = \phi(\Vect{x}) e^{ i \theta(\Vect{x}) }
\end{equation}

\noindent ultimately leads to the local dispersion relation that governs the phase function $\theta(\Vect{x})$,
\begin{subequations}
    \begin{equation}
        \Symb{D}[\Vect{x}, \nabla \theta(\Vect{x})] = 0
        ,
        \label{eq:dZERO}
    \end{equation}
    
    \noindent and the transport equation that governs the envelope function $\phi(\Vect{x})$,
    \begin{equation}
        \Vect{v}(\Vect{x})^\intercal 
        \pd{\Vect{x}} \log  \phi(\Vect{x})
        = 
        - \frac{1}{2} \pd{\Vect{x}} \cdot \Vect{v}(\Vect{x}) 
        .
        \label{eq:GOenv}
    \end{equation}
\end{subequations}

\noindent Here, we have defined the local group velocity as
\begin{equation}
    \Vect{v}(\Vect{x})
    \doteq
    \pd{\Vect{k}}
    \Symb{D}\left[\Vect{x}, \nabla \theta(\Vect{x}) \right]
    ,
\end{equation}

\noindent and we have introduced the dispersion function $\Symb{D}(\Vect{x}, \Vect{k})$, obtained from $D(\Vect{x}, \Vect{x}')$ using the Wigner transform%
~\footnote{Strictly speaking, the Wigner transform is a mapping between Hilbert-space operators and phase-space functions. In \Eqs{eq:wigner} and \eq{eq:wigFUNC}, we choose to represent the abstract operators $\hat{D}$ and $\ket{\psi} \bra{\psi}$ explicitly by their configuration-space ($\Vect{x}$-space) matrix elements for convenience.}%
\begin{equation}
    \Symb{D}(\Vect{x}, \Vect{k})
    \doteq \int \dd \Vect{s} \, 
    e^{i\Vect{k}^\intercal \Vect{s}} \,
    D
    \left(
        \Vect{x} 
        - \frac{\Vect{s}}{2}
        , 
        \Vect{x} 
        + \frac{\Vect{s}}{2}
    \right)
    .
    \label{eq:wigner}
\end{equation}

\noindent (Vectors are interpreted as row vectors unless explicitly transposed via $^\intercal$, so $\Vect{k}^\intercal \Vect{s} = \Vect{k} \cdot \Vect{s} $. Also, the symbol $\doteq$ denotes definitions.) Note that the Wigner transform of the two-point correlation function $\psi(\Vect{x})\psi^*(\Vect{x}')$, \ie
\begin{equation}
    W_\psi(\Vect{x}, \Vect{k}) \doteq 
    \int \frac{\dd \Vect{s}}{(2\pi)^N} \, 
    e^{i\Vect{k}^\intercal \Vect{s}} \,
    \psi
    \left(
        \Vect{x} 
        - \frac{\Vect{s}}{2}
    \right)
    \psi^*
    \left(
        \Vect{x} 
        + \frac{\Vect{s}}{2}
    \right)
    ,
    \label{eq:wigFUNC}
\end{equation}

\noindent acts as a phase-space (quasi-)distribution function for the field intensity, satisfying~\cite{Case08}
\begin{subequations}
    \begin{align}
        |\psi(\Vect{x})|^2 &= \int \dd \Vect{k} \, W_\psi(\Vect{x}, \Vect{k})
        , \\
        |\fourier{\psi}(\Vect{k})|^2 &= \int \dd \Vect{x} \, W_\psi(\Vect{x}, \Vect{k}),
    \end{align}
\end{subequations}

\noindent where $\fourier{\psi}$ is the Fourier transform of $\psi$.

The local dispersion relation \eq{eq:dZERO} is commonly solved via the ray equations
\begin{equation}
    \pd{\xi} \Vect{x}
    = \pd{\Vect{k}} \Symb{D}(\Vect{x}, \Vect{k})
    ,
    \quad
    \pd{\xi} \Vect{k}
    = - \pd{\Vect{x}} \Symb{D}(\Vect{x}, \Vect{k}) ,
    \label{eq:rays}
\end{equation}

\noindent where the evolution of $\Vect{k}$ along a ray (considered as a vector field over $\Vect{x}$) is constrained by
\begin{subequations}%
    \label{eq:rayCONSTR}
    \begin{equation}
        \Vect{k} = \nabla \theta(\Vect{x})
        ,
    \end{equation}
    
    \noindent and the initial conditions must satisfy
    \begin{equation}
        \Symb{D}(\Vect{x}, \Vect{k}) = 0
        .
    \end{equation}
\end{subequations} 

\begin{figure*}
    \centering
    \begin{overpic}[width = \linewidth,trim={2mm 3mm 3mm 2mm},clip]{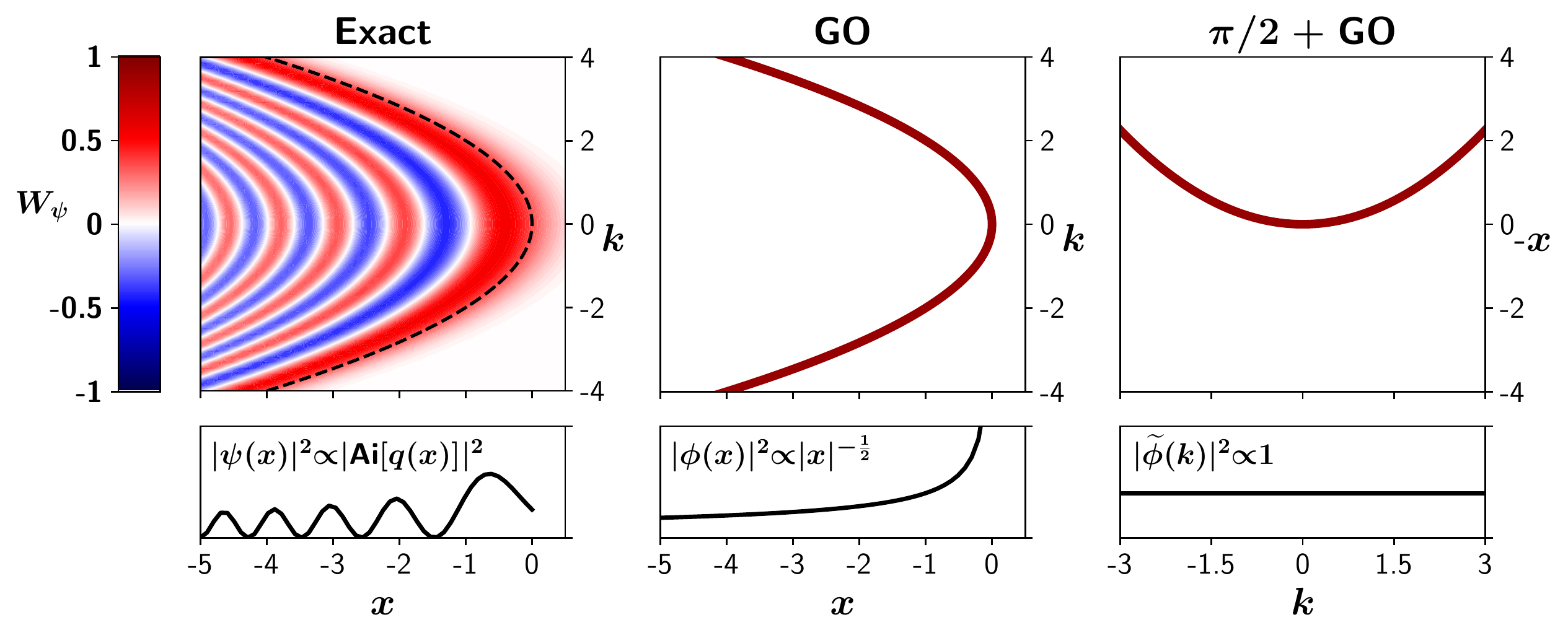}
        \put(32,16){\textbf{\large(a)}}
        \put(62,16){\textbf{\large(b)}}
        \put(92,16){\textbf{\large(c)}}
        \put(32,10){\textbf{\large(d)}}
        \put(62,10){\textbf{\large(e)}}
        \put(92,10){\textbf{\large(f)}}
    \end{overpic}
    \caption{Top row -- the phase-space distribution of the wave intensity in the phase space $(x, k)$ for a wave near a fold-type Airy caustic (see \Sec{sec:exAIRY}): \textbf{(a)} The exact Wigner function $W_\psi \propto \airyA[ \Symb{D}(x,k)]$; \textbf{(b)} GO solution $W_\psi \sim \delta\left[ \Symb{D}(x,k) \right]$; \textbf{(c)} GO solution in the phase space rotated by $\pi/2$. Bottom row -- the corresponding wave fields: \textbf{(d)} the exact field $|\psi(x)|^2$; \textbf{(e)} the field intensity $|\phi(x)|^2 \doteq \langle|\psi(x)|^2 \rangle $ in the GO approximation (where the angular brackets denote averaging over the local wavelength); \textbf{(f)} the field intensity in the spectral representation, $|\fourier{\phi}(k)|^2 \doteq \langle|\fourier{\psi}(k)|^2 \rangle$. The dispersion manifold $\Symb{D}(x,k) = 0$ is shown as the black dashed line.  Clearly, $|\phi(x)|^2$ diverges at $x\to 0$, but $|\fourier{\phi}(k)|^2$ is well-behaved everywhere.}
    \label{fig:wigner}
\end{figure*}
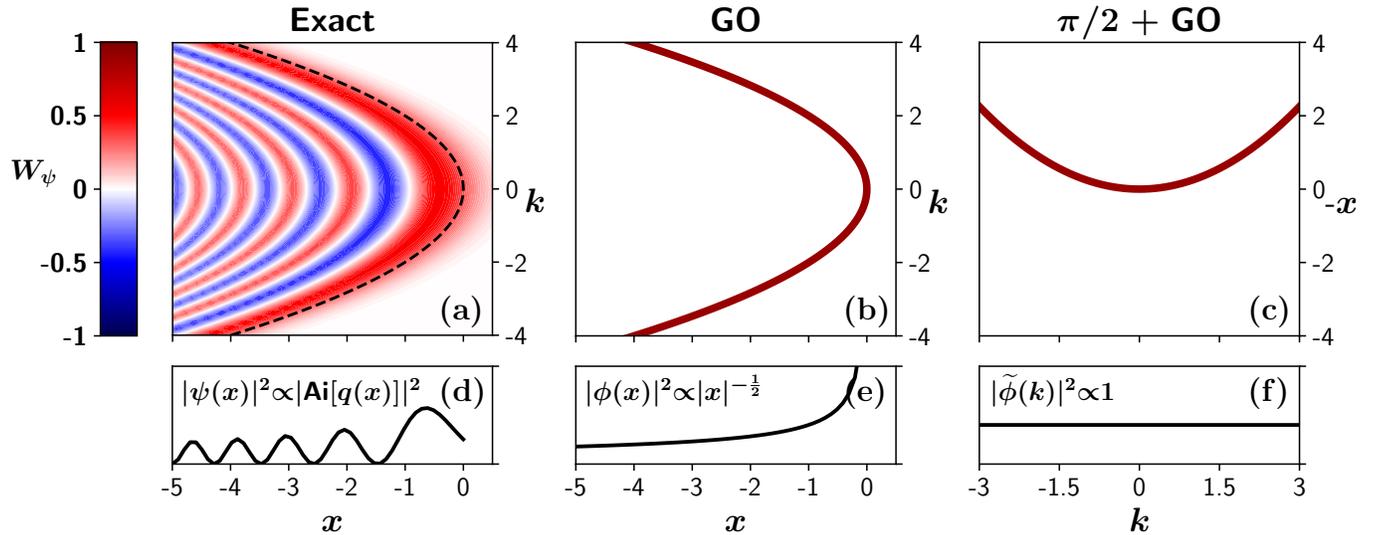

\noindent The rays encoded by \Eq{eq:rays} naturally reside in the $2N$-D ray phase space $(\Vect{x}, \Vect{k})$ with the wavevector $\Vect{k}$ serving as the ray momentum. The constraints \eq{eq:rayCONSTR} define an $N$-D manifold in this phase space called the `dispersion manifold' on which $\psi$ is asymptotically confined. More specifically, if we parameterize this manifold as the zero set of $N$ independent functions~\cite[pp.~467--468]{Tracy14} 
\begin{equation}
    \Vect{\Symb{M}}(\Vect{x}, \Vect{k}) \doteq (\Symb{M}_1(\Vect{x}, \Vect{k}), \ldots \Symb{M}_N(\Vect{x}, \Vect{k}))
    , \quad
    \Symb{M}_1 \equiv \Symb{D}
    ,
\end{equation}

\noindent then $W_\psi(\Vect{x}, \Vect{k}) \sim \delta[\Vect{\Symb{M}}(\Vect{x},\Vect{k})]$ in the GO limit~\cite{Berry77a,Berry77b}. Correspondingly, $|\phi(\Vect{x})|^2$ diverges in the GO limit (\Fig{fig:wigner}) at locations where the projection of the dispersion manifold onto $\Vect{x}$-space is singular, \ie where%
~\footnote{Note that $\pd{\Vect{k}} \Vect{\Symb{M}}\left[\Vect{x}, \nabla \theta(\Vect{x}) \right] = \pd{\Vect{\tau}} \Vect{x}(\Vect{\tau}) $ when $\Vect{\tau}$ is generated by $\Vect{\Symb{M}}$, \ie $\pd{\Vect{k}} \Symb{M}_j\left[\Vect{x}, \nabla \theta(\Vect{x}) \right] = \pd{\tau_j} \Vect{x}(\Vect{\tau})$.}%
\begin{equation}
    \det \pd{\Vect{k}} 
    \Vect{\Symb{M}}\left[\Vect{x}, \nabla \theta(\Vect{x}) \right]
    = 0,
\end{equation}

\noindent or equivalently, where
\begin{equation}
    \det \partial^2_{\Vect{x} \Vect{x}} \theta(\Vect{x}) 
    \to \infty
    .
\end{equation}

\noindent These locations are called `caustics'~\cite{Kravtsov93}, and they typically occur where distinct branches of the dispersion manifold coalesce. Notice, though, that the ray equations \eq{eq:rays} in these regions remain well-defined.

The singularity of $\phi$ at caustics signifies that the traditional GO approximation fails in regions where the projection of the dispersion manifold onto $\Vect{x}$-space is ill-behaved. However, working in $\Vect{x}$-space is not a necessity. One can instead formulate GO on more general phase-space planes specifically chosen to avoid singular projections of the dispersion manifold as the wave propagates. This is the main idea of the MGO method as proposed in \Refs{Lopez20a, Lopez21a} (see \Fig{fig:wigner}). To develop this idea in more detail, let us introduce an $N$-D coordinate system $\Vect{\tau}$ such that the dispersion manifold can be parameterized as $(\Vect{x}(\Vect{\tau}), \Vect{k}(\Vect{\tau}) )$. We choose $\tau_1 = \xi$, the longitudinal coordinate along the ray trajectories \eq{eq:rays}. The remaining $\tau_2, \ldots, \tau_N$, which can be chosen as the coordinates generated by $\Symb{M}_2, \ldots, \Symb{M}_N$, parameterize the initial conditions to \Eqs{eq:rays} along the dispersion manifold, \eg the remaining spatial coordinates.

Let us consider a specific point $\Vect{\tau} = \Vect{t}$ on the dispersion manifold. The optimal rotation to avoid projection singularities is the one that aligns the tangent plane of the dispersion manifold at $\Vect{t}$, denoted $\Vect{X}_\Vect{t}$-space, with $\Vect{x}$-space. This is accomplished by performing the following linear coordinate transformation of the phase space:
\begin{equation}
    \begin{pmatrix}
        \Vect{X}_\Vect{t} \\
        \Vect{K}_\Vect{t}
    \end{pmatrix}
    =
    \Mat{S}_\Vect{t}
    \begin{pmatrix}
        \Vect{x} \\
        \Vect{k}
    \end{pmatrix}
    ,
    \quad
    \Mat{S}_\Vect{t}
    \doteq
    \begin{pmatrix}
        \Mat{A}_\Vect{t} & \Mat{B}_\Vect{t} \\
        \Mat{C}_\Vect{t} & \Mat{D}_\Vect{t}
    \end{pmatrix}
    ,
    \label{eq:rotPHASE}
\end{equation}

\noindent where $\Mat{S}_\Vect{t}$ is the $2N \times 2N$ unitary symplectic matrix that rotates phase space to align $\Vect{X}_\Vect{t}$-space with $\Vect{x}$-space. (The matrices $\Mat{A}_\Vect{t}$, $\Mat{B}_\Vect{t}$, $\Mat{C}_\Vect{t}$, and $\Mat{D}_\Vect{t}$ are each $N \times N$.) An explicit construction of $\Mat{S}_\Vect{t}$ using ray trajectories near $\Vect{t}$ is provided by the `symplectic Gram--Schmidt' algorithm of \Ref{Lopez20a}.

Ultimately, repeating these `optimal' rotations for all points on the dispersion manifold, synthesizing the resulting GO solutions, and transforming them back to the original phase space yields the MGO solution~\cite{Lopez20a}
\begin{equation}
    \psi(\Vect{x})
    =
    \sum_{\Vect{t} \in \Vect{\tau}(\Vect{x})} 
    \MTnorm(\Vect{x}) \,
    \Upsilon_\Vect{t}(\Vect{x})
    ,
    \label{eq:MGO}
\end{equation}

\noindent where $\Vect{\tau}(\Vect{x})$ is the function inverse of $\Vect{x}(\Vect{\tau})$; accordingly, the sum is taken over all branches of $\Vect{k}(\Vect{x}) \doteq \Vect{k}[\Vect{\tau}(\Vect{x})]$. In \Eq{eq:MGO}, we have introduced the integral function
\begin{align}
    \Upsilon_\Vect{t}(\Vect{x})
    &\doteq 
    \int_{\cont{0}} \dd \Vect{\varepsilon} \,
    \Psi_\Vect{t}\left[ 
        \Vect{\varepsilon} + \Vect{X}_\Vect{t}(\Vect{t})
    \right] 
    \exp\left(
        - \frac{i}{2} \Vect{\varepsilon}^\intercal \Mat{D}_\Vect{t} \Mat{B}_\Vect{t}^{-1} \Vect{\varepsilon}
    \right)
    \nonumber\\
    &\hspace{15mm}\times
    \exp\left\{
        i
        \Vect{\varepsilon}^\intercal
        \Mat{B}_\Vect{t}^{-\intercal}
        \left[ 
            \Vect{x}
            -
            \Mat{D}_\Vect{t}^\intercal \Vect{X}_\Vect{t}(\Vect{t})
        \right]
    \right\}
    ,
    \label{eq:upsilon}
\end{align}

\noindent where $\Psi_\Vect{t}(\Vect{X}_\Vect{t})$ is the GO solution in the rotated phase space \eq{eq:rotPHASE}. The integration is performed along the steepest-descent contour that passes through the saddlepoint $\Vect{\varepsilon} = \Vect{0}$ (\Sec{sec:steepDESC}), which importantly means that $\Upsilon_\Vect{t}(\Vect{x})$ is not generally a unitary mapping of $\Psi_\Vect{t}$ unless $\cont{0}$ can be deformed to lie along the real axis%
~\footnote{The fact that $\Upsilon_\Vect{t}(\Vect{x})$ is not generally unitary does not greatly diminish the accuracy of MGO, since the multiple contributing $\Upsilon_\Vect{t}(\Vect{x})$ are summed in such a manner to keep the solution finite [\Eq{eq:MGO}].}. %
Note that \Eq{eq:upsilon} requires $\Mat{B}_\Vect{t}$ to be invertible; this is done for simplicity, and the generalization to arbitrary $\Mat{B}_\Vect{t}$ is provided in \Ref{Lopez21a}. Also, the prefactor $\MTnorm(\Vect{x})$ in \Eq{eq:MGO} can be simply evolved along the ray trajectories using the formulas provided in \Ref{Lopez20a}. Efficiently computing $\Upsilon_\Vect{t}(\Vect{x})$ is comparatively less understood, and shall be the focus of the remainder of this work.


\subsection{Caustics as optical catastrophes}
\label{sec:caustic}

\begin{table*}
    \centering
    \begin{tabular}{| c | c | c | c | c |}
        \multicolumn{1}{c}{\textbf{Name}} & \multicolumn{1}{c}{$\alpha$} & \multicolumn{1}{c}{$m$} & \multicolumn{1}{c}{$M$} & \multicolumn{1}{c}{$f_\alpha(\Vect{\kappa}, \Vect{y})$} \\
        \hline
        No caustic & $A_1$ & $0$ & $1$ 
        & $\kappa_1^2$ \\
        Fold & $A_2$ & $1$ & $1$ 
        & $\kappa_1^3 + y_1 \kappa_1$ \\
        Cusp & $A_3$ & $2$ & $1$ 
        & $\kappa_1^4 + y_2 \kappa_1^2 + y_1 \kappa_1$ \\
        Swallowtail & $A_4$ & $3$ & $1$ 
        & $\kappa_1^5 + y_3 \kappa_1^3 + y_2 \kappa_1^2 + y_1 \kappa_1$ \\
        Hyperbolic umbilic & $D_4^+$ & $3$ & $2$ 
        & $\kappa_1^3 + \kappa_2^3 + y_3 \kappa_1 \kappa_2 + y_2 \kappa_2 + y_1 \kappa_1$ \\[1mm]
        Elliptic umbilic & $D_4^-$ & $3$ & $2$ 
        & $\kappa_1^3 - 3 \kappa_1 \kappa_2^2 + y_3(\kappa_1^2 + \kappa_2^2) + y_2 \kappa_2 + y_1 \kappa_1$ \\[1mm]
        \hline
    \end{tabular}
    \caption{A complete list of the normal-form generators $f_\alpha(\Vect{\kappa}, \Vect{y})$ for caustics with codimension $m \le 3$~\cite{Berry80b, Arnold83, Kravtsov93}. The interference patterns that correspond to these caustics can be viewed in their entirety when the number of spatial dimensions $N = 3$. For each caustic, $\alpha$ is the Arnold label~\cite{Arnold83}, $m$ is the codimension, and $M$ is the corank.}
    \label{tab:caustic}
\end{table*}

The behavior of $\Upsilon_\Vect{t}(\Vect{x})$ near caustics can be broadly understood using catastrophe theory~\cite{Berry80b,Kravtsov93, Poston96}. This theory provides a classification system for caustics based on their codimension, that is, the minimum number of spatial dimensions in which they can be observed. For example, the simplest type of caustic is the fold caustic, which occurs when a wave encounters a cutoff. The fold caustic has codimension $1$, so it can be observed in $N$-D systems with $N \ge 1$. On the other hand, the cusp caustic, which occurs at a focal point, has codimension $2$ and can thus only be observed for $N \ge 2$. This supports the intuition that cutoffs are well-described by $1$-D models like Airy's equation, but foci are inherently $2$-D.

There are three main advantages to using the catastrophe classification system to study caustics: \textbf{(i)} Only `structurally stable' caustics that are robust under small perturbations are included. These are the caustics that are most physically relevant, since a structurally unstable caustic will be destroyed by any imperfections in the experimental setup (which are of course unavoidable). For example, an EM wave propagating in an unmagnetized cold plasma with a linear density profile $n(x)$ will have a cutoff at some $x_c$. If the density is perturbed from $n(x)$ to $\tilde{n}(x)$ by some global motion of the plasma, the cutoff location will shift from $x_c$ to $\tilde{x}_c$, but will generally not disappear; hence, the cutoff (fold caustic) is `structurally stable'. \textbf{(ii)} There are only a finite number of distinct caustic types that are stable in a given number of dimensions. For example, only six different caustics can occur in $3$-D (including `no caustic'; see Table~\ref{tab:caustic}). \textbf{(iii)} General properties of a given caustic type can be determined by studying a single member in detail, often chosen to be the `simplest' member (the so-called `normal-form generator' of the caustic class; see below).

These three results from catastrophe theory greatly reduce the work required to validate any new method in catastrophe optics; indeed, a new method for modeling caustics need only be tested on three different caustics to be fully viable in $2$-D, or on six different caustics for $3$-D. In our case, a numerical quadrature rule for $\Upsilon_\Vect{t}(\Vect{x})$ can be validated on the standard integrals of catastrophe theory, which are integral representations for caustic wavefields and take the general form
\begin{equation}
    I_\alpha(\Vect{y}) \doteq 
    \int \dd \Vect{\kappa} \,
    \exp\left[
        i f_\alpha(\Vect{\kappa}, \Vect{y})
    \right]
    ,
    \label{eq:Ialpha}
\end{equation}

\noindent where $\Vect{y}$ is an $m$-D collection of `external' (or `control') variables, $\Vect{\kappa}$ is an $M$-D collection of `internal' (or `state') variables, and $\alpha$ labels the type of caustic. The integers $m \le N$ and $M \le N$ are the `codimension' and `corank' of the caustic, respectively, and the function $f_\alpha(\Vect{\kappa}, \Vect{y})$ is the normal-form generator for a type-$\alpha$ caustic. For example, the fold caustic, also called the $A_2$ caustic in Arnold's nomenclature~\cite{Arnold83}, has $m = 1$, $M = 1$, and
\begin{equation}
    f_{A_2}(\kappa_1, y_1)
    = \kappa_1^3 + y_1 \kappa_1
    .
\end{equation}

\noindent The corresponding $I_{A_2}(y_1)$ is proportional to the Airy function $\airyA(y_1/ \sqrt[3]{3})$~\cite[pp.~194--213]{Olver10a}. See Table~\ref{tab:caustic} for more examples. Note that if $M < N$ such that only a subset of the integration variables of $\Upsilon_\Vect{t}$ are included in the standard integral $I_\alpha$, then by the splitting lemma of catastrophe theory~\cite{Berry80b,Poston96}, the remaining $N - M$ integrals contained in $\Upsilon_\Vect{t}$ are decoupled and involve phase functions that are quadratic at most, and thereby trivially integrated. 

For practical problems, $I_\alpha(\Vect{y})$ typically represents only the local behavior of a given type-$\alpha$ caustic. The global behavior can sometimes be modeled using the method of `uniform approximation'~\cite{Olver10a,Chester57,Ludwig66}. However, this method relies on \textbf{(i)} the caustic type being known beforehand (which is fine for interpretive but not predictive simulations) and \textbf{(ii)} only a single caustic being present. Indeed, the elementary catastrophes mentioned here often combine to form `caustic networks', an example being an EM wave focused on a cutoff producing a fold-cusp network. It might be possible to infer basic properties of such caustic networks from the constituent members, but complete understanding can only be achieved by considering the caustic network as a whole, which is very difficult to do analytically. Hence, a robust numerical scheme for computing catastrophe integrals that does not assume any specific caustic structure is needed.


\section{Gauss--Freud quadrature for steepest-descent integration}

\label{sec:algorithm}


\subsection{Steepest-descent method}
\label{sec:steepDESC}

As seen from \Eq{eq:Ialpha} and Table~\ref{tab:caustic}, we generally expect $I_\alpha(\Vect{y})$ to involve a highly oscillatory integrand when $\Vect{y}$ and $\Vect{\kappa}$ are both real. These rapid oscillations would make the direct evaluation of $\Upsilon_\Vect{t}(\Vect{x})$ analytically and numerically challenging, if not for the fact that $\Upsilon_\Vect{t}(\Vect{x})$ is evaluated along the steepest-descent contour $\cont{0}$ [\Eq{eq:upsilon}]. Along $\cont{0}$, oscillatory terms become exponentially decaying terms, reinstating the viability of standard numerical integration methods like Gaussian quadrature (\Sec{sec:gaussQUAD}). However, this simplification is contingent on the ability to determine $\cont{0}$ for arbitrary wavefields. Let us therefore characterize the steepest-descent contours of the standard forms $I_\alpha(\Vect{y})$. For simplicity, we restrict attention to $1$-D integrals, \ie $M = 1$ in \Eq{eq:Ialpha}.

For integrals of the form
\begin{equation}
    I(\Vect{y}) = \int \dd \kappa \, g(\kappa, \Vect{y}) \exp[i f(\kappa, \Vect{y}) ]
    \label{eq:SDMint}
\end{equation}

\noindent [where we have generalized \Eq{eq:Ialpha} to include a slowly varying amplitude $g(\kappa, \Vect{y})$], the steepest-descent contours at fixed $\Vect{y}$ are by definition the streamlines of $\nabla \Im(f)$ in the complex $\kappa$ plane $\mbb{C}^1$, where $\nabla \doteq (\pd{\Re(\kappa)},  \pd{\Im(\kappa)} )$. (Here, $\Re$ and $\Im$ denote the real and imaginary parts of a complex function.) When $f$ is analytic in $\kappa$, a more useful definition arises from the Cauchy--Riemann relation~\cite{Rudin87equation}
\begin{equation}
    i \pd{\Re(\kappa)} f(\kappa, \Vect{y}) = \pd{\Im(\kappa)} f(\kappa, \Vect{y})
    ,
    \label{eq:cauchy1}
\end{equation}

\noindent which implies that $\nabla \Re(f)$ and $\nabla \Im(f)$ are orthogonal, \ie
\begin{equation}
    \nabla \Re(f) \cdot \nabla \Im(f) = 0
    .
    \label{eq:cauchy2}
\end{equation}

\noindent Therefore, the streamlines of $\nabla \Im(f)$ that pass through a given point $\kappa_0$ are also the set of points in $\mbb{C}^1$ that satisfy the implicit equation
\begin{equation}
    \Re\left[ f(\kappa, \Vect{y}) \right] = \Re\left[ f(\kappa_0, \Vect{y}) \right].
\end{equation}

The steepest-descent contours are almost-everywhere smooth curves, although non-differentiable kinks can occur at special points where \Eq{eq:cauchy2} is indeterminate, \ie where $\nabla \Re\left(f \right) = \nabla \Im\left(f \right) = \Vect{0}$%
~\footnote{By \Eq{eq:cauchy1}, $\nabla \Re\left(f \right)$ and $\nabla \Im\left(f \right)$ always vanish simultaneously.}. %
At these points [which are saddlepoints per \Eq{eq:cauchy1}%
~\footnote{This follows by differentiating \Eq{eq:cauchy1} to show that the Hessian matrices for $\Re(f)$ and $\Im(f)$ are symmetric and traceless, thereby possessing two real eigenvalues of opposite sign.}%
], the direction of $\nabla \Im\left(f \right)$ generally changes abruptly, producing the aforementioned kinks that require special parameterization. As shown later, such parameterization can be done by treating the kink as two independent curves that intersect at a finite angle.

Although saddlepoints are `rare' in that they occur at isolated points in $\mbb{C}^1$, they are often of primary interest due to their prominent role in asymptotic wave theory. More specifically, each saddlepoint of $f$ encodes the contribution to $I(\Vect{y})$ from a single corresponding GO ray. Consequently, a saddlepoint $\kappa_s(\Vect{y})$ will be real when $\Vect{y}$ is in the lit region of a caustic, but may be complex when $\Vect{y}$ is in the shadow region. Also, a caustic occurs at the specific values of $\Vect{y}$, denoted $\Vect{y}_c$, such that multiple saddlepoints coalesce and consequently, 
\begin{equation}
    \pd{\kappa}^2 f[\kappa_s(\Vect{y}_c), \Vect{y}_c] = 0
    .
    \label{eq:caustic}
\end{equation}

Having now characterized the general behavior of steepest-descent contours, in the following subsections, we shall briefly overview the Gaussian quadrature method~\cite{Press07gauss}, and then show how it can be used to accurately compute $\Upsilon_\Vect{t}(\Vect{x})$ along $\cont{0}$.


\subsection{Gaussian quadrature for numerical integration }
\label{sec:gaussQUAD}

Suppose we wish to compute the integral of some real-valued function $h(\kappa)$ over the real interval $(a,b)$, with both $a$ and $b$ allowed to be infinite. Suppose further that $h(\kappa)$ can be partitioned as 
\begin{equation}
    h(\kappa) = \omega(\kappa) r(\kappa)
    ,
    \label{eq:GQdecomp}
\end{equation} 

\noindent with $\omega(\kappa)$ positive-definite on $(a,b)$ and $r(\kappa)$ a polynomial of degree $2n - 1$. Then, the following $n$-point quadrature formula holds:
\begin{equation}
    \int_a^b \dd \kappa \, h(\kappa)
    \equiv
    \int_a^b \dd \kappa \, \omega(\kappa) r(\kappa)
    = \sum_{j = 1}^n w_j r(\kappa_j)
    ,
    \label{eq:gqEXACT}
\end{equation}

\noindent where the quadrature weights $\{ w_j \}$ and nodes $\{ \kappa_j \}$ are determined as follows. 

Let us introduce the inner product
\begin{equation}
    \langle h_1, h_2 \rangle 
    \doteq \int_a^b \dd \kappa \, \omega(\kappa) h_1(\kappa) h_2(\kappa)
    .
    \label{eq:innerPROD}
\end{equation}

\noindent Let us also introduce the family of real-valued polynomials $\{p_\ell(\kappa)\}$ (with $\ell$ the polynomial degree) that are orthogonal with respect to \Eq{eq:innerPROD}, that is,
\begin{equation}
    \langle p_\ell, p_m \rangle
    = \eta_\ell \, \delta_{\ell m}
    , \quad
    \eta_\ell \doteq \langle p_\ell, p_{\ell} \rangle
    ,
\end{equation}

\noindent where $\delta_{\ell m}$ is the Kronecker delta. By performing polynomial division of $r(\kappa)$ by $p_n(\kappa)$ and Lagrange interpolation of the residual, it can be shown~\cite[pp.~135--137]{Gil07} that the quadrature weights $\{w_j \}$ are determined by the formula
\begin{equation}
     w_j = 
     \left\langle \prod_{\substack{\ell = 1 \\ j \neq \ell}}^n
     \frac{\kappa - \kappa_\ell}{\kappa_j - \kappa_\ell}, 1 \right\rangle
     = \frac{\langle p_n(\kappa), (\kappa - \kappa_j)^{-1} \rangle}{p'_n(\kappa_j)}
     ,
     \label{eq:GQweights}
\end{equation}

\noindent and the quadrature nodes are the $n$ zeros of $p_n(\kappa)$, \ie
\begin{equation}
    \{\kappa_j\} = \{ \kappa ~ | ~ p_n(\kappa) = 0 \}
    .
    \label{eq:GQnodes}
\end{equation}

If $h(\kappa)$ cannot be decomposed as \Eq{eq:GQdecomp} with polynomial $r(\kappa)$, the corresponding integral can still be approximately computed as
\begin{equation}
    \int_a^b \dd \kappa \, h(\kappa)
    \approx
    \sum_{j = 1}^n w_j \frac{h(\kappa_j)}{\omega(\kappa_j)}
    .
    \label{eq:gq}
\end{equation}

\noindent Equation \eq{eq:gq} defines the Gaussian quadrature method of numerical integration. The error in using \Eq{eq:gq} depends on how `close' $h(\kappa)/\omega(\kappa)$ is to being a $2n-1$ degree polynomial, as determined by the maximum value of $\pd{\kappa}^{2n}(h/\omega)$ over $(a,b)$. Explicitly,~\cite{Suli03equation}
\begin{equation}
    \left|
        \int_a^b \dd \kappa \, h(\kappa)
        -
        \sum_{j = 1}^n w_j \frac{h(\kappa_j)}{\omega(\kappa_j)}
    \right|
    \le
    \frac{\eta_n}{(2n)!}
    \max_{\zeta \in (a,b)} 
    \left|
        \pd{\kappa}^{2n} \frac{h(\zeta)}{\omega(\zeta)}
    \right|
    .
    \label{eq:gqERROR}
\end{equation}

\noindent Note that the right-hand side vanishes when $h(\kappa)/\omega(\kappa)$ is a $2n-1$ degree polynomial, as desired. Also note that \Eq{eq:gq} is still valid when $h(\kappa)$ is complex-valued.

Common choices for $\{p_\ell(\kappa) \}$ are the rescaled Legendre polynomials for integrals over finite $(a,b)$ with $\omega(\kappa) = 1$, or the Hermite polynomials for integrals with $(a,b) = (-\infty, +\infty)$ and $\omega(\kappa) = \exp(-\kappa^2)$. For our purposes, though, we will find it more convenient to use the less-common Freud polynomials (\App{sec:FreudQUAD}), as we shall now explain.


\subsection{Gauss--Freud quadrature}

Let us now develop the appropriate Gaussian quadrature rule for MGO. Along the steepest-descent contour that passes through a given saddlepoint at $\kappa = \kappa_0$, denoted $\cont{0}$, \Eq{eq:SDMint} takes the general form
\begin{align}
    I(\Vect{y}_0) = 
    &\exp 
    \left\{
        i \Re[f(\kappa_0, \Vect{y}_0)] 
        \nullFrac
    \right\}
    \nonumber\\
    &\times
    \int_{\cont{0}} \dd \kappa \, 
    g(\kappa, \Vect{y}_0)
    \exp
    \left\{
        - \Im
        \left[
            f(\kappa, \Vect{y}_0) 
            \nullFrac
        \right]
    \right\}
    ,
\end{align}
    
\noindent or equivalently,
\begin{align}
    I(\Vect{y}_0) = 
    &\exp 
    \left\{
        i \Re[f(\kappa_0, \Vect{y}_0)] 
        \nullFrac
    \right\}
    \nonumber\\
    &\times
    \int_{-\infty}^\infty \dd l \,
    \kappa'(l)
    g[\kappa(l), \Vect{y}_0]
    \exp
    \left[
        - F(l,\Vect{y}_0)
    \right]
    ,
    \label{eq:steepI}
\end{align}

\noindent where we have introduced $\kappa(l)$ as a $1$-D parameterization of $\cont{0}$ with $\kappa(0) = \kappa_0$, we have defined
\begin{equation}
    F(l, \Vect{y}_0) \doteq 
    \Im
    \left\{
        f[\kappa(l), \Vect{y}_0] 
        \nullFrac
    \right\}
    ,
\end{equation}

\noindent and we have set $\Vect{y} = \Vect{y}_0$ to emphasize that $\Vect{y}$ should be considered a \textit{fixed parameter} for the integration over $\kappa$%
~\footnote{Specifically, $\Vect{y}$ is related to the physical location of the wavefield $\psi(\Vect{x})$ via the ray map $\Vect{x}(\Vect{\tau})$ along with the local coordinate transformation $\Vect{\tau}(\Vect{y})$ needed to place $\Upsilon_\Vect{t}(\Vect{x})$ into standard form.}. %
Note that $\kappa_0$ being a saddlepoint implies
\begin{equation}
    \nabla \Im
    \left[
        f(\kappa_0, \Vect{y}_0) 
        \nullFrac
    \right]
    = \Vect{0}
    , \quad
    \pd{l} F(0,\Vect{y}_0)
    = 0 ,
    \label{eq:saddleI}
\end{equation}

\noindent and $\mc{C}_0$ being a steepest-descent contour implies
\begin{equation}
    F(l, \Vect{y}_0) \ge F(0, \Vect{y}_0)
    .
    \label{eq:Fdecrease}
\end{equation}

Suppose first that $\kappa_0$ is a non-degenerate saddlepoint, which typically occurs when $\Vect{y}_0$ does not coincide with a caustic. This means that
\begin{equation}
    \pd{l}^2 F(0, \Vect{y}_0) > 0
    .
    \label{eq:nonDEGEN}
\end{equation}

\noindent Hence, $F(l, \Vect{y}_0)$ is well-approximated around $l = 0$ as
\begin{equation}
    F(l, \Vect{y}_0) \approx F(0, \Vect{y}_0) + \pd{l}^2 F(0, \Vect{y}_0) \, l^2
    .
    \label{eq:taylorF}
\end{equation}

\begin{figure*}
    \centering
    \includegraphics[width=0.32\linewidth,trim={16mm 4mm 26mm 4mm},clip]{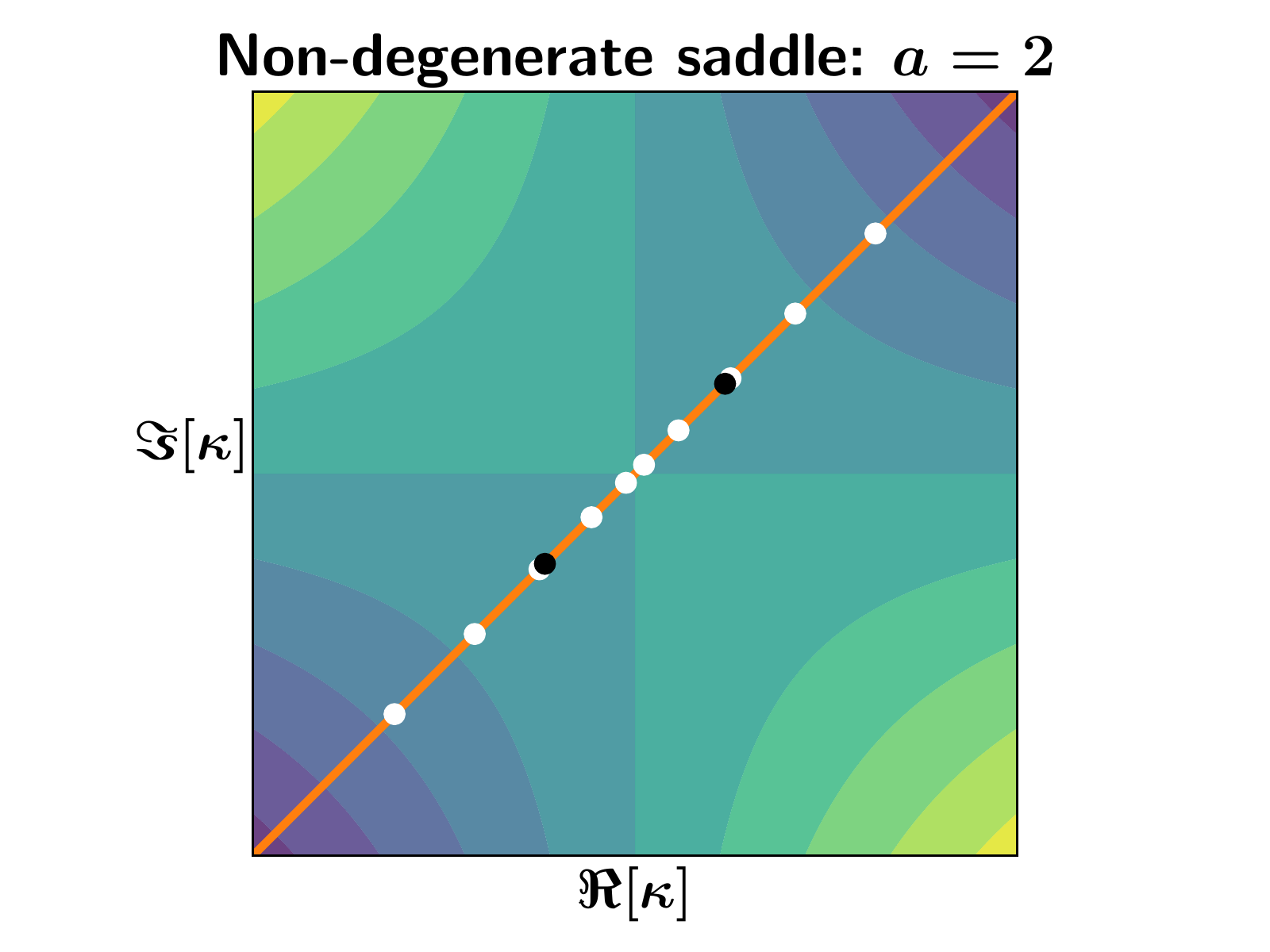}
    \includegraphics[width=0.32\linewidth,trim={16mm 4mm 26mm 4mm},clip]{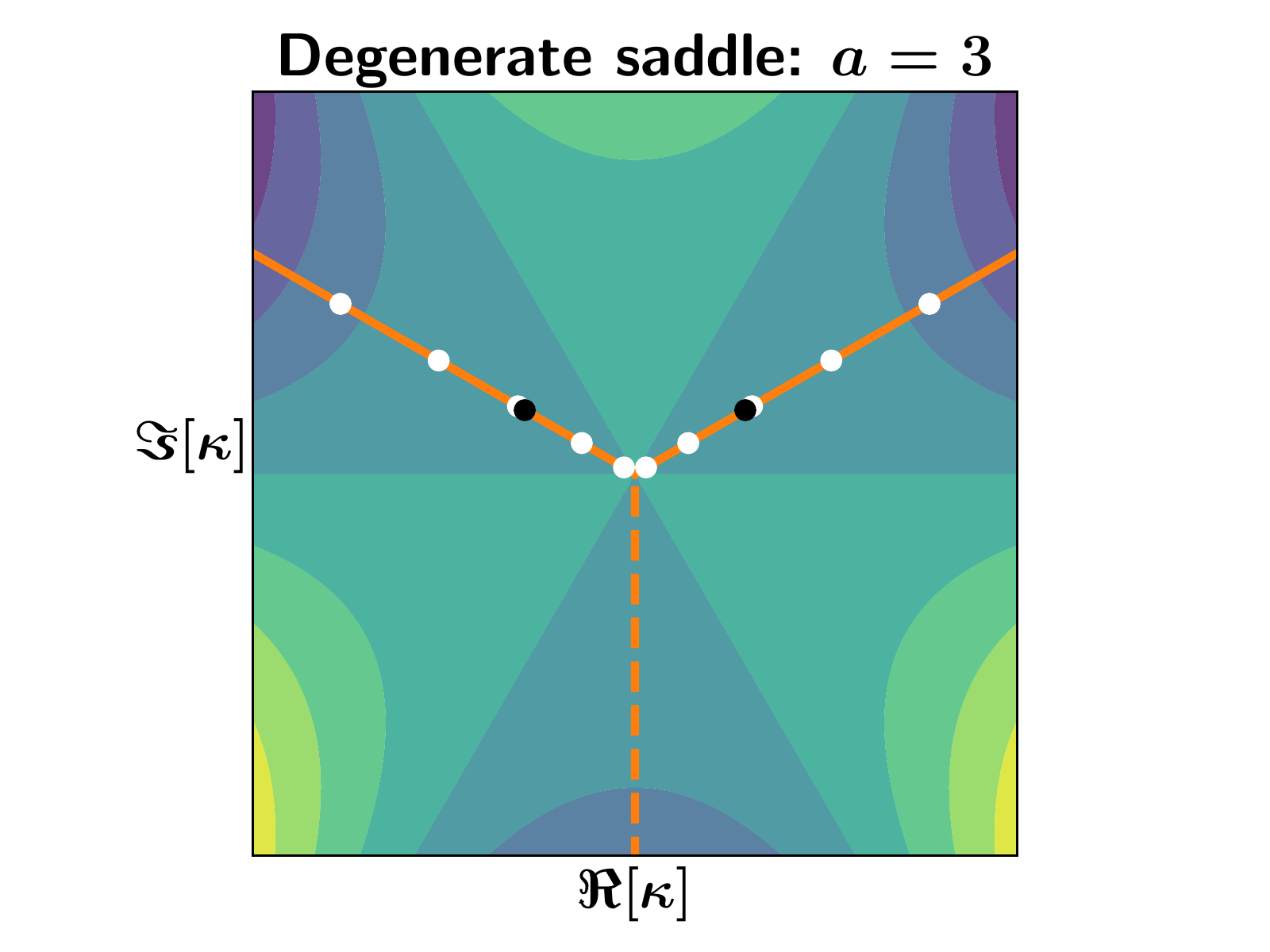}
    \includegraphics[width=0.32\linewidth,trim={16mm 4mm 26mm 4mm},clip]{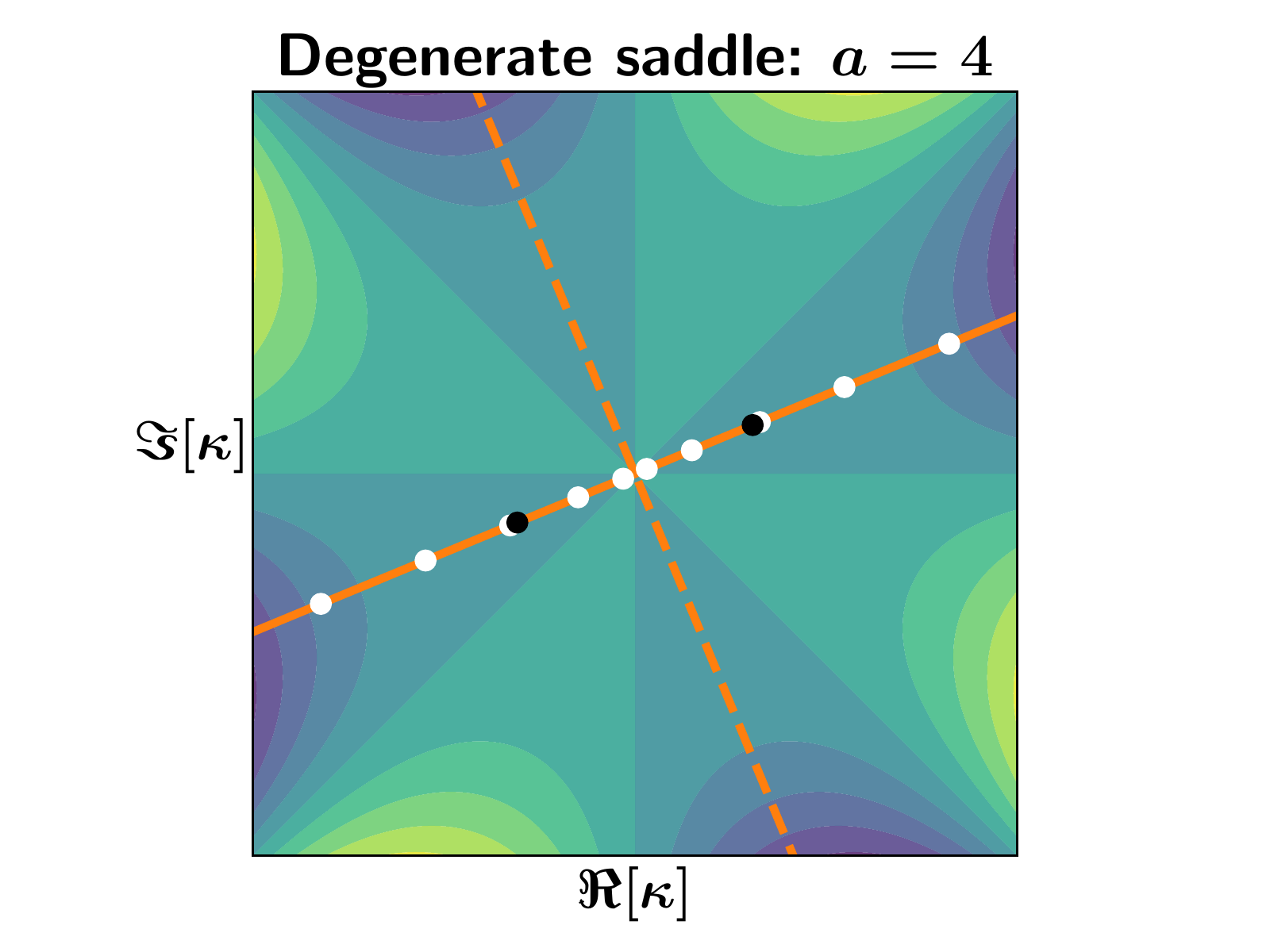}
    \caption{Steepest-descent contours (orange) for the integrand phase function $f(\kappa) \doteq \kappa^a$ of \Eq{eq:EXgauss} for various values of the parameter $a$, which characterizes the saddlepoint degeneracy. The background color represents the magnitude of the integrand $-\Im(f)$, with green corresponding to larger values and blue corresponding to smaller values. The order $n = 5$ quadrature nodes are shown as white dots, while the points $\kappa(l_\pm)$ [\Eq{eq:kappaL}] that are used to determine the rotation angles $\sigma_\pm$ [\Eq{eq:sigmaDEF}] are shown as black dots. Unused steepest-descent contours are shown as dashed orange lines. As can be seen, the steepest-descent contour has a kink when $a$ is odd, which requires $\sigma_+$ and $\sigma_-$ to be calculated separately.}
    \label{fig:EXcontour}
\end{figure*}

\noindent However, this approximation has obvious issues when $\kappa_0$ is a degenerate saddlepoint, which occurs when $\Vect{y}_0$ coincides with a caustic. For this case, although by \Eq{eq:caustic},
\begin{equation}
    \pd{l}^2 F(0, \Vect{y}_0) = 0
    ,
    \label{eq:degen}
\end{equation}

\noindent a quadratic function can still be fit to $F(l, \Vect{y}_0)$ as 
\begin{equation}
    F(l, \Vect{y}_0) \approx F(0, \Vect{y}_0) + s(\Vect{y}_0) \, l^2
    ,
    \label{eq:secantF}
\end{equation}

\noindent provided the scaling factor $s(\Vect{y}_0)$ is chosen appropriately; we choose to use the finite-difference formula
\begin{subequations}
    \begin{align}
        \label{eq:sDEF}
        s(\Vect{y}_0) 
        &= 
        \left\{
            \begin{array}{lr}
                s_-(\Vect{y}_0), & l \le 0 \\
                s_+(\Vect{y}_0), & l > 0
            \end{array}
        \right.
        , \\
        \label{eq:sPM}
        s_\pm(\Vect{y}_0) 
        &\doteq
        \frac{F(l_\pm, \Vect{y}_0) - F(0, \Vect{y}_0)}{ | l_\pm |^2 }
        ,
    \end{align}
\end{subequations}

\noindent where $l_\pm$ satisfy the threshold condition
\begin{equation}
    F(l_\pm, \Vect{y}_0) - F(0, \Vect{y}_0) \ge C_\pm 
    \label{eq:kappaL}
\end{equation}

\noindent with $C_\pm$ arbitrary constants. (We use $C_\pm = 1$ for simplicity, but we found varying $C_\pm$ over the range $[0.5, 5]$ produced only $O(10^{-5})$ differences.) Note that we have allowed the possibility for different scaling factors on either side of $l = 0$ in case $\cont{0}$ has a kink at $\kappa_0$ (see \Fig{fig:EXcontour}). Importantly, \Eq{eq:Fdecrease} implies that $s > 0$. Also note that \Eq{eq:secantF} reduces to \Eq{eq:taylorF} in the limit $C_\pm \to 0$ when $\kappa_0$ is non-degenerate.

As a final simplification, let us adopt a piecewise linear approximation to $\cont{0}$ such that%
~\footnote{The mapping $\kappa(l)$ being nonlinear is not a problem for Gaussian quadrature \textit{per se}, but it necessitates the inclusion of expensive root-finding steps into the Gaussian quadrature algorithm~\cite{Deano09}}%
\begin{equation}
    \kappa(l) \approx \kappa_0 + |l| \times
    \left\{
        \begin{array}{lr}
            \exp(i \sigma_-), & l \le 0 \\
            \exp(i \sigma_+), & l > 0
        \end{array}
    \right.
    \label{eq:unionC0}
\end{equation}

\noindent for suitable rotation angles $\sigma_\pm$, which are allowed to be different in case $\cont{0}$ has a kink at $\kappa_0$. It would be natural to choose $\sigma_\pm$ such that \Eq{eq:unionC0} is a tangent-line approximation to $\cont{0}$ at $\kappa_0$; however, this choice cannot be applied to degenerate saddlepoints. Instead, we shall allow \Eq{eq:unionC0} to generally describe the secant-line approximations to $\cont{0}$ that underlie \Eq{eq:sPM}, namely,
\begin{align}
    \sigma_\pm 
    &= \text{arg}
    \left[
        \kappa(l_\pm) - \kappa_0
    \right]
    \nonumber\\
    &=
    \text{sign}
    \left\{
        \frac{
            \Im\left[
                \kappa(l_\pm) - \kappa_0
            \right]
        }{
            \| \kappa(l_\pm) - \kappa_0 \|
        }
    \right\}
    \cos^{-1}
    \left\{
        \frac{
            \Re\left[
                \kappa(l_\pm) - \kappa_0
            \right]
        }{
            \| \kappa(l_\pm) - \kappa_0 \|
        }
    \right\}
    \label{eq:sigmaDEF}
    .
\end{align}

\noindent Here, we take the convention that $\text{sign}(0) = 1$; hence \Eq{eq:sigmaDEF} restricts $\sigma_\pm$ to lie on the interval $(-\pi, \pi]$.

Inserting \Eqs{eq:secantF}, \eq{eq:sDEF}, and \eq{eq:unionC0} into \Eq{eq:steepI} yields
\begin{widetext}
\begin{align}
    \frac{
        I(\Vect{y}_0)
    }{
        \exp[i f(\kappa_0,\Vect{y}_0)]
    }
    &\approx
    \int_0^\infty \dd l 
    \left\{
        g[\kappa_0 + l \exp(i \sigma_+), \Vect{y}_0]
        \exp[i \sigma_+ - s_+(\Vect{y}_0) l^2]
        -
        g[\kappa_0 + l \exp(i \sigma_-), \Vect{y}_0]
        \exp[i \sigma_- - s_-(\Vect{y}_0) l^2]
        \nullFrac
    \right\}
    \nonumber\\
    &=
    \int_0^\infty \dd l 
    \left\{
        g
        \left[
            \kappa_0 
            + \frac{ l \exp(i \sigma_+) }{ \sqrt{s_+(\Vect{y}_0)} }
            , \Vect{y}_0
        \right]
        \frac{
            \exp(i \sigma_+)
        }{
            \sqrt{s_+(\Vect{y}_0)} 
        }
        -
        g
        \left[
            \kappa_0 
            + \frac{ l \exp(i \sigma_-) }{ \sqrt{s_-(\Vect{y}_0)} }
            , \Vect{y}_0
        \right]
        \frac{
            \exp(i \sigma_-)
        }{
            \sqrt{s_-(\Vect{y}_0)} 
        }
    \right\}
    \exp(- l^2)
    .
\end{align}

\noindent Hence, an appropriate Gaussian quadrature rule for MGO is based on the inner product
\begin{equation}
    \langle h_1, h_2 \rangle
    = \int_0^\infty \dd l \, h_1(l) h_2(l) 
    \exp(-l^2)
    ,
    \label{eq:FreudIP}
\end{equation}

\noindent for which Freud polynomials are orthogonal (\App{sec:FreudQUAD}). Using \Eq{eq:gq}, we obtain the quadrature rule for MGO:
\begin{equation}
    I(\Vect{y}_0)
    \approx 
    \sum_{j = 1}^n 
    w_j \exp(l_j^2)
    \left\{
        h\left[
            \kappa_0 
            + \frac{l_j \exp(i \sigma_+)}{\sqrt{s_+(\Vect{y}_0)}}, \Vect{y}_0
        \right]
        \frac{\exp(i \sigma_+)}{\sqrt{s_+(\Vect{y}_0)}}
        -
        h\left[
            \kappa_0 
            + \frac{l_j \exp(i \sigma_-)}{\sqrt{s_-(\Vect{y}_0)}}, \Vect{y}_0
        \right]
        \frac{\exp(i \sigma_-)}{\sqrt{s_-(\Vect{y}_0)}}
    \right\}
    ,
    \label{eq:MGOquad}
\end{equation}
\end{widetext}

\noindent where $h(\kappa) = g(\kappa, \Vect{y}_0) \exp[i f(\kappa, \Vect{y}_0) ]$, and $\{l_j \}$ are the quadrature nodes. [We use $\{ l_j \}$ rather than $\{ \kappa_j \}$ to be consistent with the notation of \Eq{eq:FreudIP}.] Since Gauss--Freud quadrature is somewhat uncommon, a table of the corresponding $\{w_j\}$ and $\{l_j\}$ for various values of $n$ is also provided in \App{sec:FreudQUAD}.


\subsection{Angle memory feedback for MGO simulations}
\label{sec:feedback}

Let us now allow $\Vect{y}$ to vary in \Eq{eq:MGOquad} (rather than being fixed at some $\Vect{y} = \Vect{y}_0$), as will occur when MGO is used to simulate a propagating wave. The steepest-descent topology for $I(\Vect{y})$ will be different for each new value of $\Vect{y}$, with correspondingly new values of $\sigma_\pm$. Repeatedly searching for $\cont{0}$ to compute $\sigma_\pm$ via \Eq{eq:sigmaDEF} can be computationally expensive, and merely identifying the correct $\cont{0}$ can be difficult in situations where multiple valid steepest-descent lines exist, as occurs at caustics. Fortunately, the steepest-descent topology of $I(\Vect{y})$ typically evolves smoothly with $\Vect{y}$, which means successive calculations of $\sigma_\pm$ will be correlated. We use this fact to construct a `memory feedback' algorithm to both speed up the time required to calculate the steepest-descent topology of $I(\Vect{y})$ and to correctly identify $\cont{0}$ at caustics.

First, let us initialize the MGO simulation far from a caustic such that $\cont{0}$ is sufficiently simple: we expect the initial angles $\sigma_\pm^{(0)}$ to be approximately given as
\begin{equation}
    \sigma_\pm^{(0)}
    \approx -\frac{\pi}{4} - \frac{\text{arg}[\pd{\kappa}^2f(\kappa_0, \Vect{y}_0)]}{2} \pm \frac{\pi}{2}
\end{equation}

\noindent restricted to the interval $(-\pi, \pi]$. By starting the search for the exact $\cont{0}$ near this value of $\sigma_\pm^{(0)}$, the search time can be reduced. As the simulation progresses, $\cont{0}$ will evolve smoothly; at each new point $\Vect{y}_j$, the search-time for $\cont{0}$ can be reduced by initializing the search near the previously calculated $\sigma_\pm^{(j-1)}$ corresponding to the previous position $\Vect{y}_{j-1}$. Moreover, by restricting the search to \textit{only} consider angles near $\sigma_\pm^{(j-1)}$, \ie restricting $|\sigma_\pm^{(j)} - \sigma_\pm^{(j-i)}| \le \Delta $ for some threshold $\Delta$ (we choose $\Delta = 0.01$), the correct $\cont{0}$ will naturally be identified by analytic continuation, even at caustics.


\section{Benchmarking results}

\label{sec:examples}


\subsection{Isolated saddlepoint}

As a first benchmarking of our numerical steepest-descent algorithm \eq{eq:MGOquad}, let us consider the numerical evaluation of the following family of integrals:
\begin{align}
    I(a,b) \doteq
    \int_{-\infty}^\infty \dd \kappa \, \kappa^b \exp\left(i \kappa^a\right)
    ,
    \quad
    a \ge 2, \,
    b \ge 0
    ,
    \label{eq:EXgauss}
\end{align}

\noindent whose exact solution is given by
\begin{equation}
    I_\textrm{ex}(a,b)
    =
    \frac{2}{a} \, \Gamma\left( \frac{2 \chi}{\pi} \right)
    \times
    \left\{
        \begin{array}{lr}
            \exp\left( i \chi \right), & \bar{a} = 0, \, \bar{b} = 0\\[1mm]
            0, & \bar{a} = 0, \, \bar{b} = 1\\[1mm]
             \cos \chi, & \bar{a} = 1, \, \bar{b} = 0\\[1mm]
             i \sin \chi, & \bar{a} = 1, \, \bar{b} = 1
        \end{array}
    \right.
    ,
\end{equation}

\noindent where we have defined
\begin{equation}
    \chi \doteq \frac{1+b}{2a} \pi
    ,
    \quad
    \bar{a} \doteq \textrm{mod}_2(a) 
    ,
    \quad
    \bar{b} \doteq \textrm{mod}_2(b) 
    .
\end{equation}

\noindent The family $I(a,b)$ also corresponds to the $A_{a-1}$ `cuspoid' caustic family (Table~\ref{tab:caustic}) evaluated at $\Vect{y} = 0$; as such, the integrand of $I(a,b)$ has an isolated saddlepoint at $\kappa = 0$ whose degeneracy is controlled by the value of $a$, with $a = 2$ being non-degenerate. To evaluate $I(a,b)$ via \Eq{eq:MGOquad} the scaling factors $s_\pm$ and rotation angles $\sigma_\pm$ are needed; these are given respectively as $s_\pm = 1$ and
\begin{equation}
    \sigma_+ = \frac{\pi}{2a}
    , \quad
    \sigma_- = 
    \left\{
        \begin{array}{lr}
            \sigma_+ - \pi, & \bar{a} = 0 \\
            \pi - \sigma_+, & \bar{a} = 1
        \end{array}
    \right.
    .
    \label{eq:EXangles}
\end{equation}

\noindent In particular, \Eq{eq:EXangles} implies that the steepest-descent contour has a kink at $\kappa = 0$ when $a$ is odd, which necessitates our partitioning of \Eq{eq:MGOquad} into incoming and outgoing branches. This feature is also shown in \Fig{fig:EXcontour}.

\begin{figure}
    \centering
    \includegraphics[width=\linewidth,trim={2mm 6mm 4mm 4mm},clip]{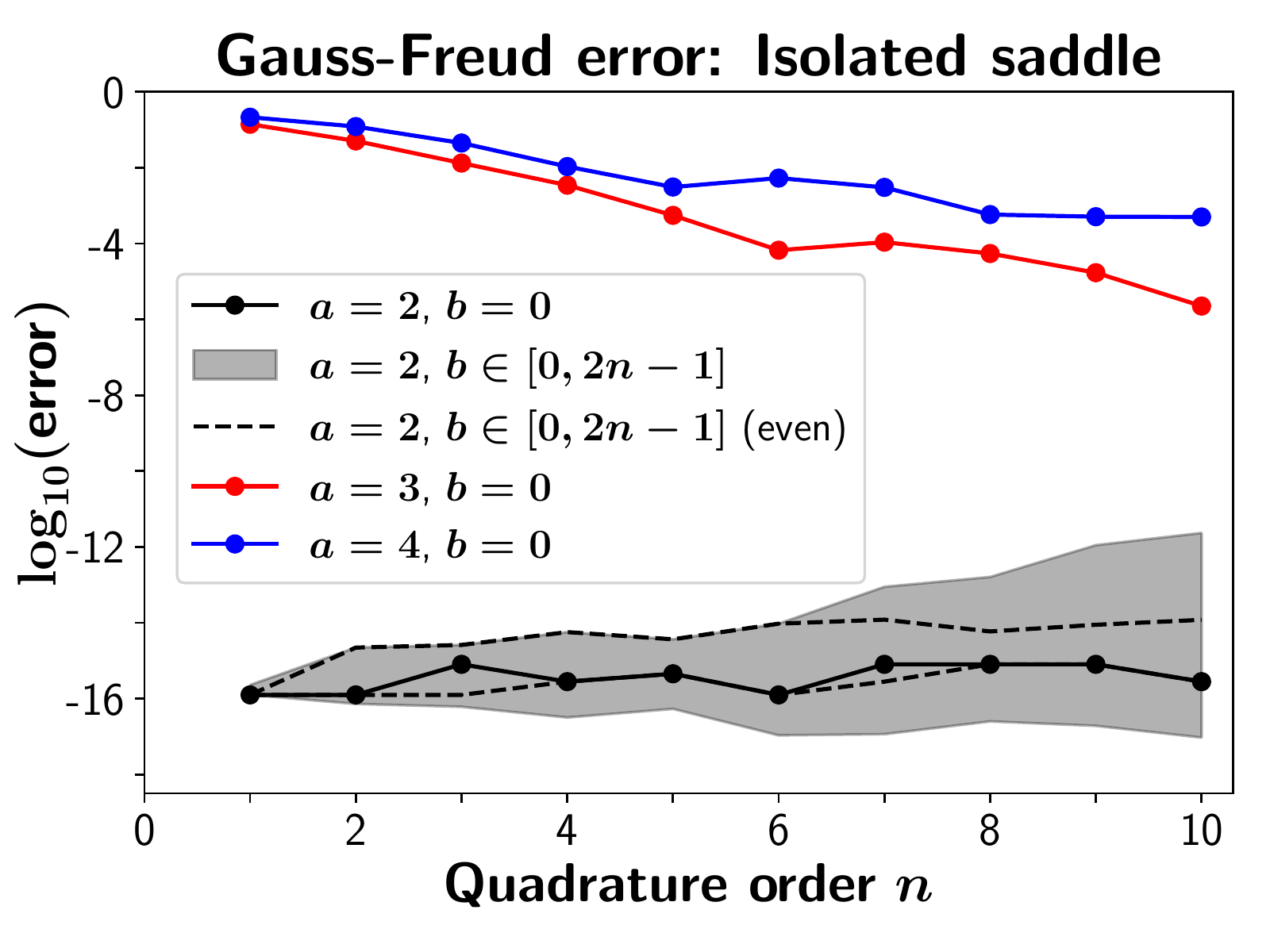}
    \caption{Comparison of the error in computing \Eq{eq:EXgauss} using the quadrature rule of \Eq{eq:MGOquad} for various values of $a$ and $b$. The error metric used is the relative error when $I(a, b)$ [Eq. (48)] is nonzero and the absolute error otherwise. The shaded gray region marks the range of error over the entire range $b \in [0, 2n-1]$ for which our quadrature rule is expected to be exact, while the dashed black lines bound the region obtained when only even values of $b$ are considered. The precision of the quadrature nodes and weights used is $10^{-15}$. }
    \label{fig:EXerr}
\end{figure}

Figure~\ref{fig:EXerr} shows the error that results from evaluating $I(a,b)$ via \Eq{eq:MGOquad} for quadrature order $n \le 10$. The quadrature weights and nodes used have precision $10^{-15}$ and are listed explicitly in Table~\ref{tab:GFnodes}. Note that our quadrature rule was developed to evaluate $I(a,b)$ exactly when $a = 2$ and $b \in [0, 2n - 1]$, and indeed, we observe that the error for these values of $a$ and $b$ remains on the order of the node/weight precision until $n = 6$, beyond which the error is slightly larger than expected. However, this increased error is not due to issues with our quadrature rule \textit{per se}, but rather due to the round-off error that unavoidably accumulates when subtracting large numbers. This conclusion is corroborated by the fact that the increased error is isolated to the cases when $b$ is even and $I(a,b)$ should be identically zero in exact arithmetic by (anti-) symmetry. When $a > 2$, our quadrature rule achieves a respectable accuracy of $10^{-4}$ even at the relatively low quadrature order of $n = 10$, demonstrating the utility of \Eq{eq:MGOquad} at caustics and regular points alike.

\subsection{EM wave in unmagnetized plasma slab with linear density profile}
\label{sec:exAIRY}

\begin{figure*}
    \centering
    \includegraphics[width=0.24\linewidth,trim={18mm 4mm 32mm 4mm},clip]{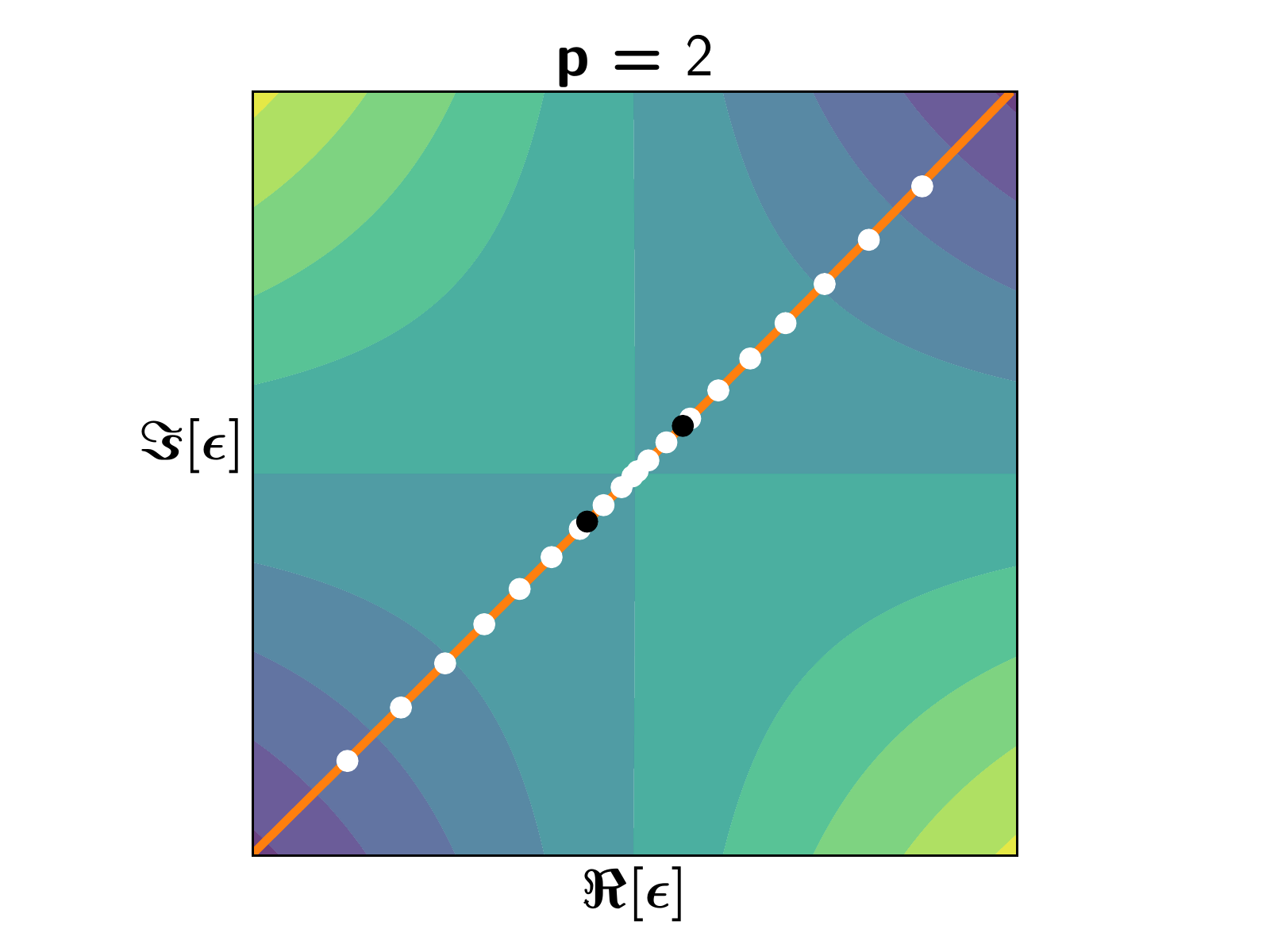}
    \includegraphics[width=0.24\linewidth,trim={18mm 4mm 32mm 4mm},clip]{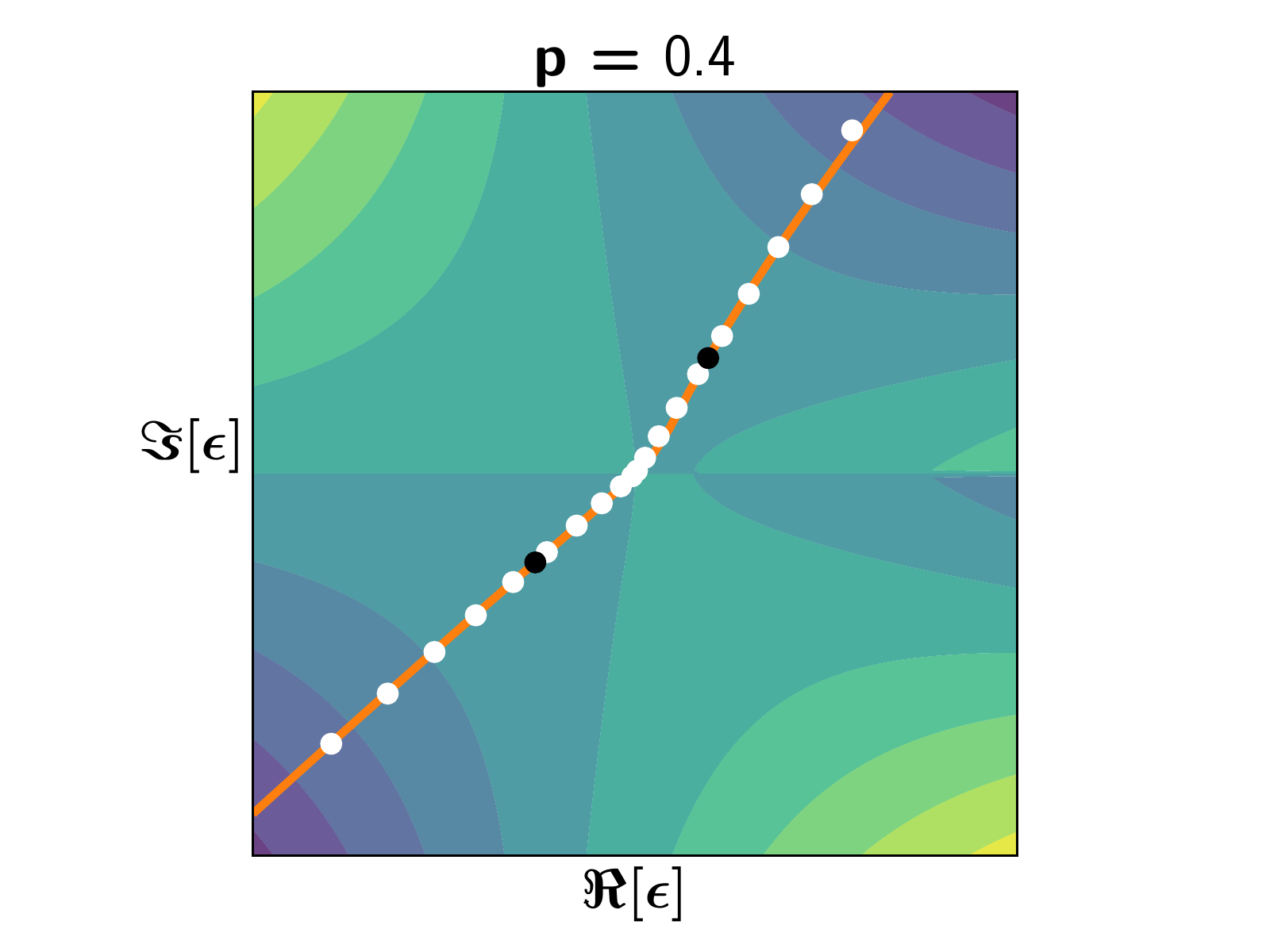}
    \includegraphics[width=0.24\linewidth,trim={18mm 4mm 32mm 4mm},clip]{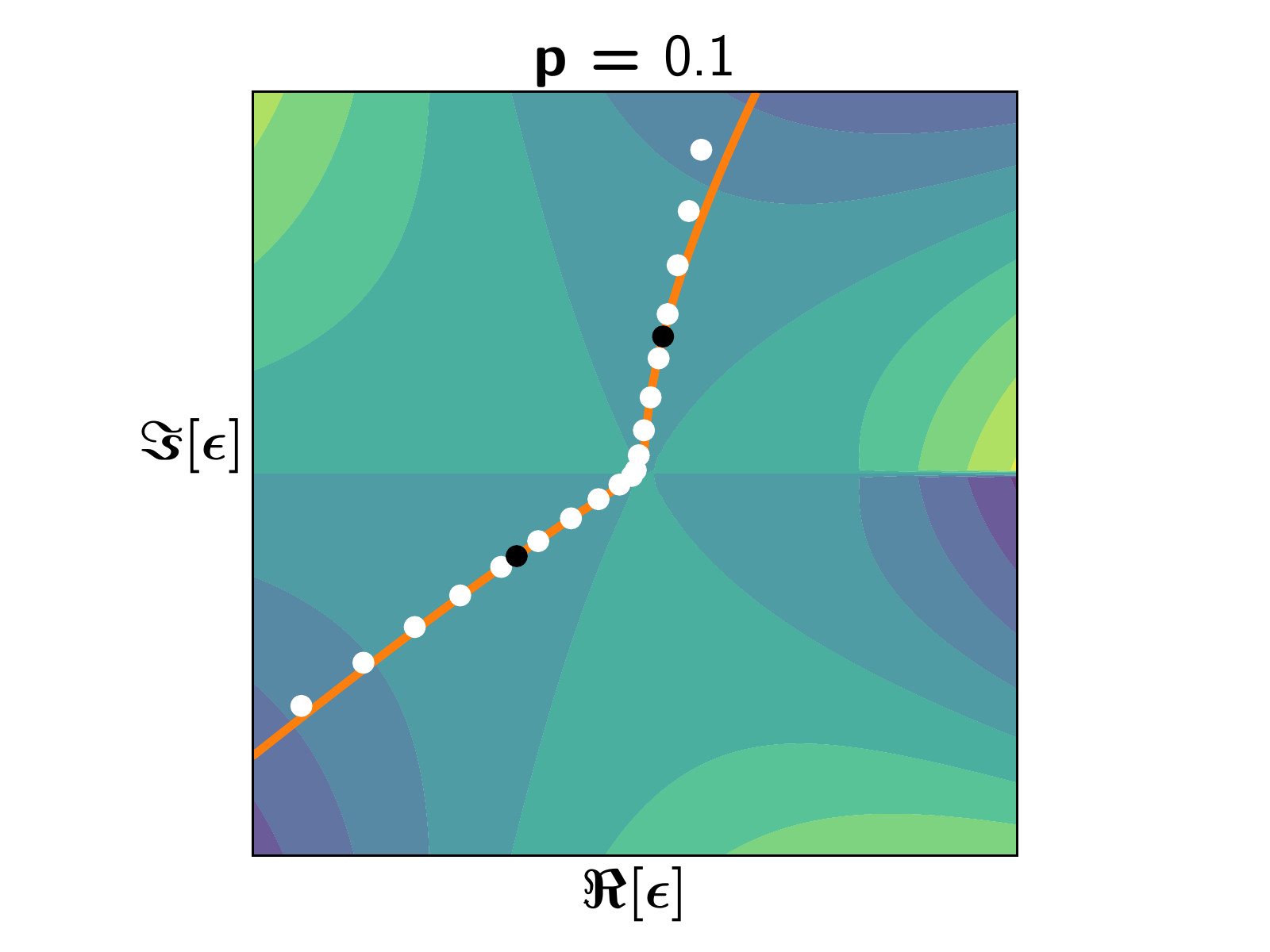}
    \includegraphics[width=0.24\linewidth,trim={18mm 4mm 32mm 4mm},clip]{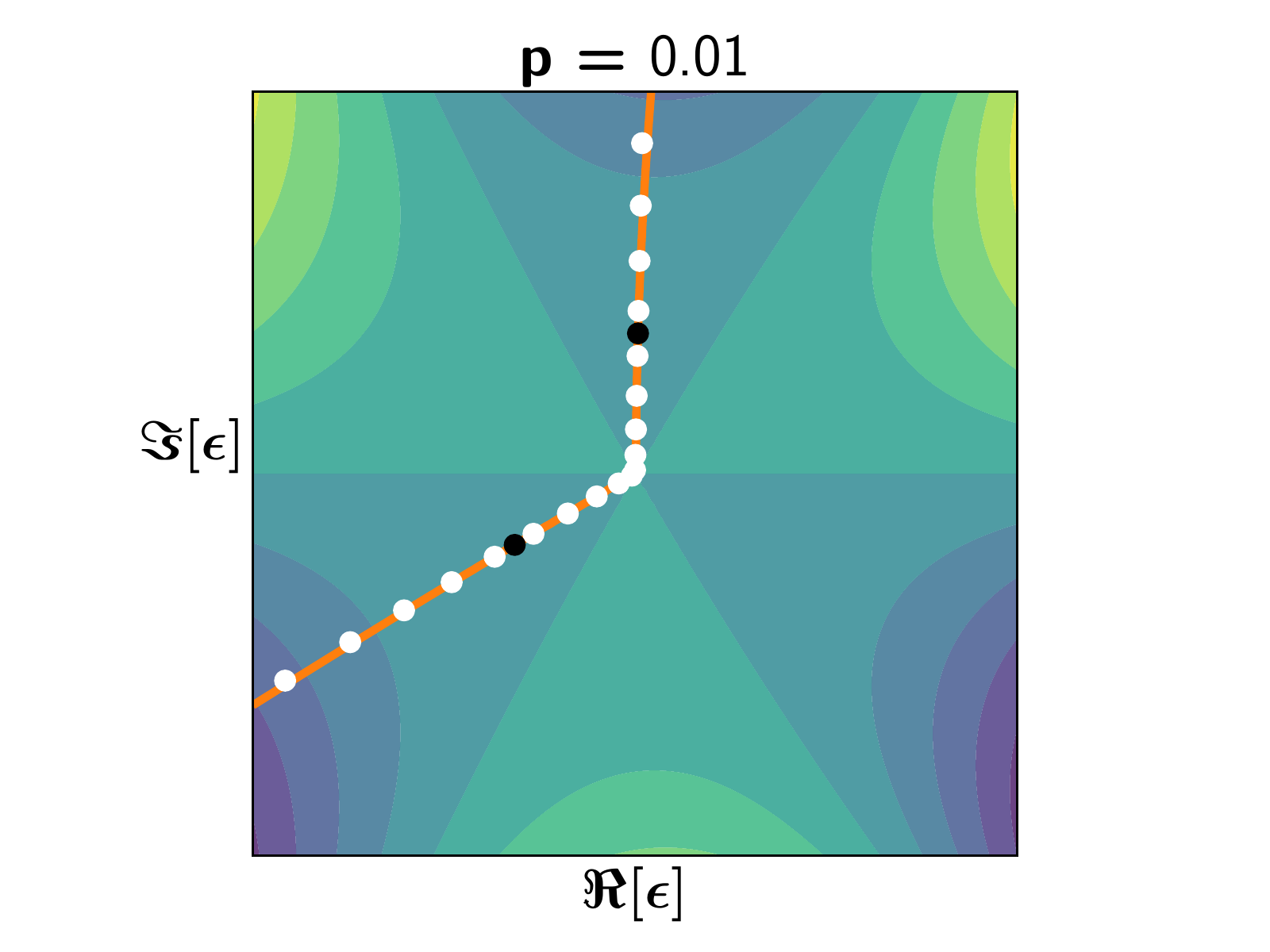}
    
    \vspace{2mm}
    \includegraphics[width=0.24\linewidth,trim={18mm 4mm 32mm 4mm},clip]{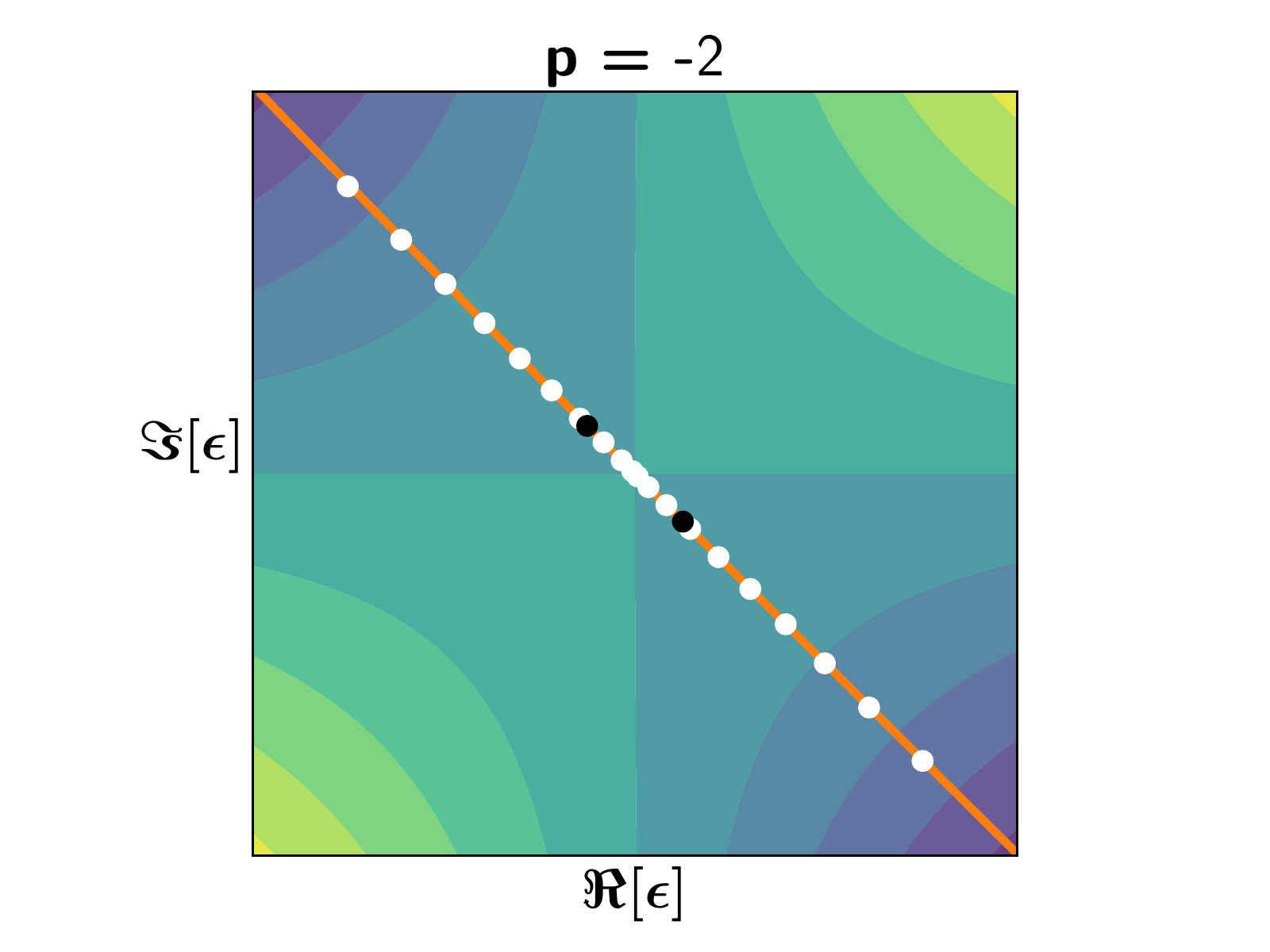}
    \includegraphics[width=0.24\linewidth,trim={18mm 4mm 32mm 4mm},clip]{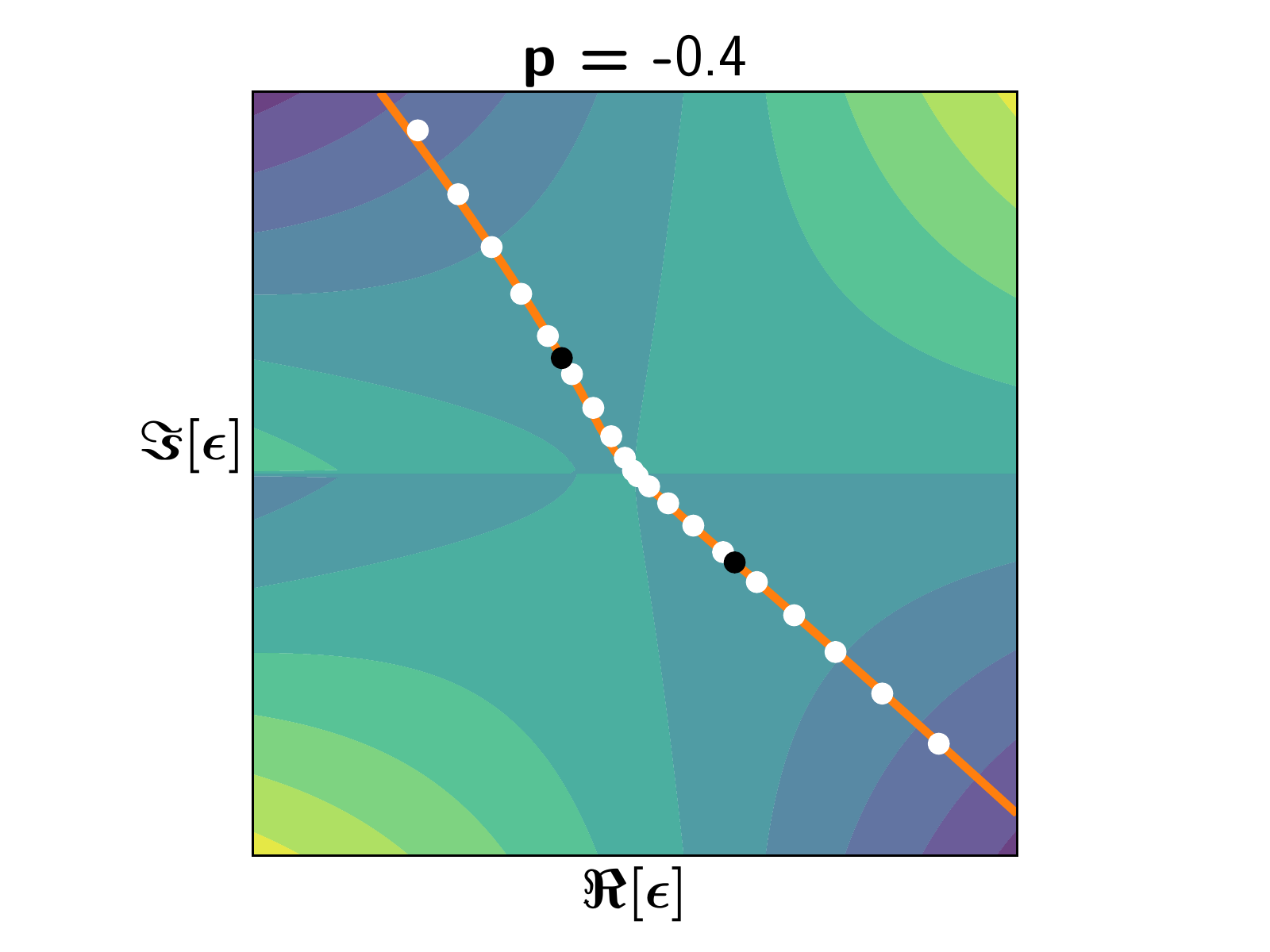}
    \includegraphics[width=0.24\linewidth,trim={18mm 4mm 32mm 4mm},clip]{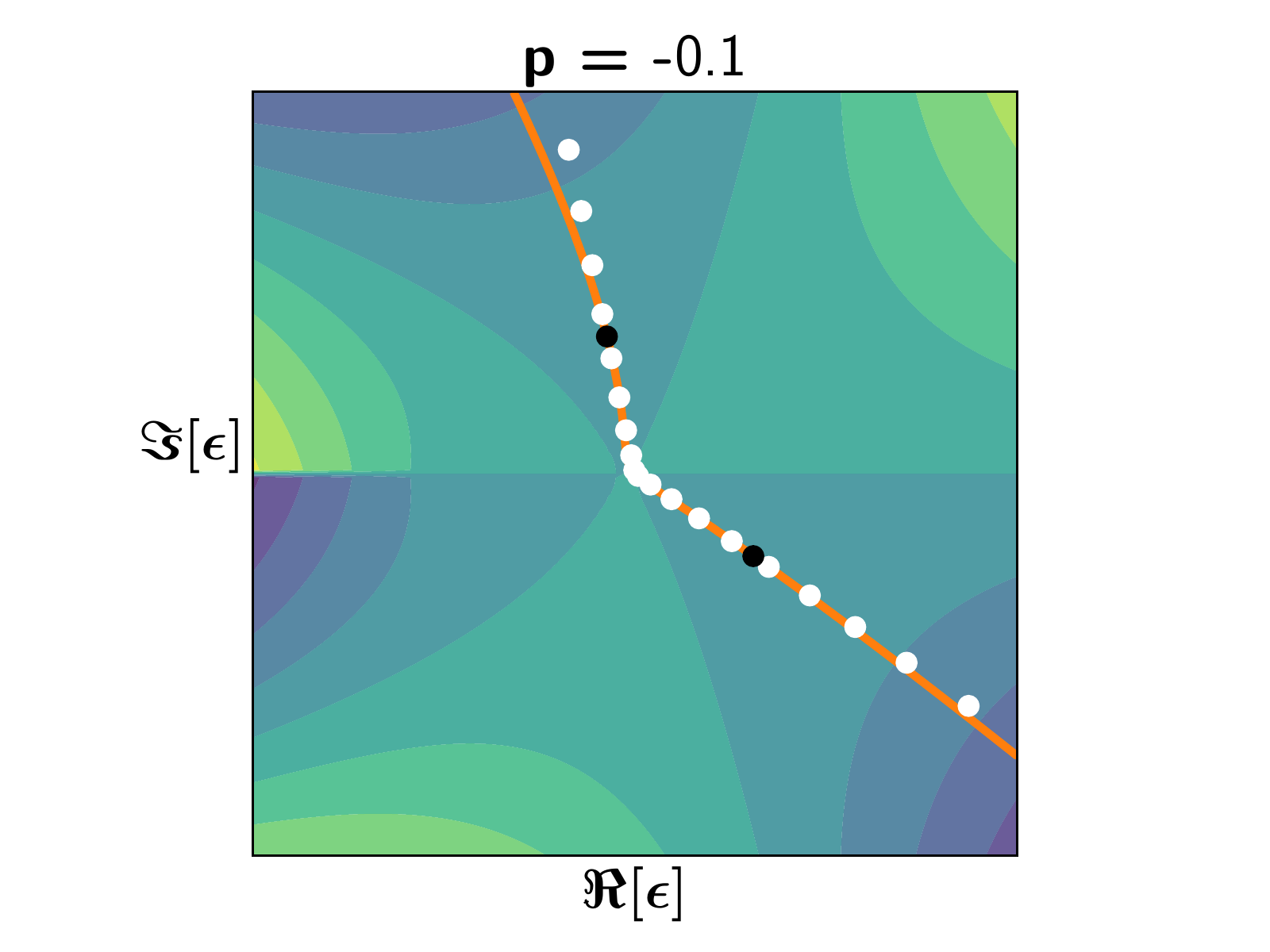}
    \includegraphics[width=0.24\linewidth,trim={18mm 4mm 32mm 4mm},clip]{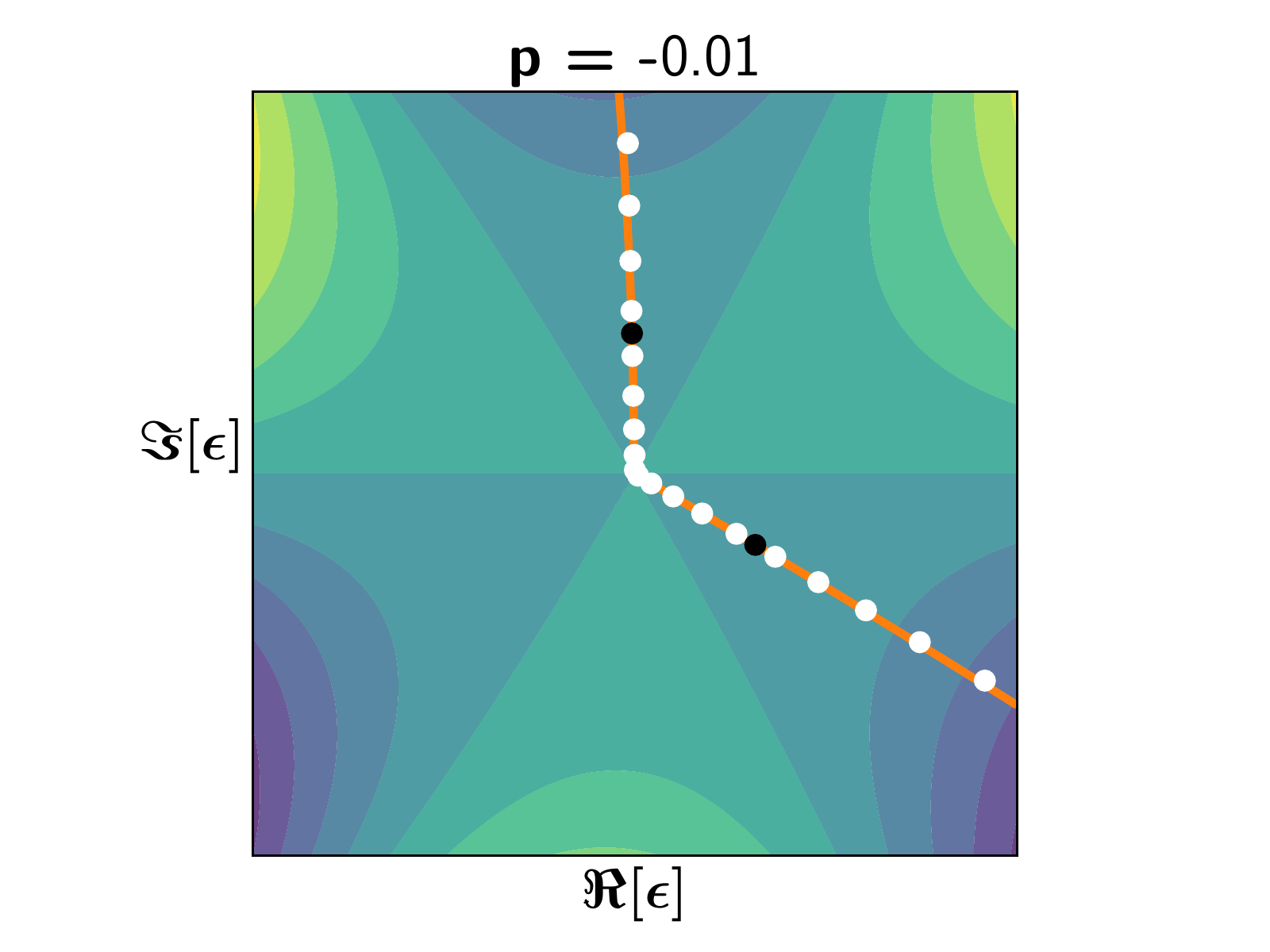}
    \caption{Same as \Fig{fig:EXcontour} for the phase function $f(\epsilon, p)$ [\Eq{eq:fAIRY}] at various values of $p$. The white dots correspond here to the $n = 10$ quadrature nodes. The steepest-descent contours evolve smoothly with $p$ and ultimately coalesce into a fold-type $A_2$ caustic at $p = 0$.}
    \label{fig:MGOcontour}
\end{figure*}

As a more realistic example, let us consider the MGO description of an EM wave propagating in a stationary unmagnetized plasma slab with a linearly varying density profile. Suppose that the EM wave and all subsequently induced fluctuations have time dependence of the form $\exp(-i \Omega t)$, where $\Omega$ is the wave frequency. Then, after defining $x$ as the direction of inhomogeneity, the electric field of the EM wave can be shown to satisfy~\cite[p.~344]{Stix92}
\begin{equation}
    \pd{x}^2 E(x) + \frac{\Omega^2}{c^2}\left[1 - \frac{n(x)}{n_c} \right] E(x) = 0
    ,
    \label{eq:waveEQ}
\end{equation}

\noindent where $c$ is the speed of light in vacuum and $n_c$ is the cutoff density. Let us assume
\begin{equation}
    n(x) = n_c \left(1 + \frac{x}{L_n} \right)
    ,
\end{equation}

\noindent where $L_n$ is some constant length scale. Then, \Eq{eq:waveEQ} takes the form
\begin{equation}
    \pd{q}^2 E(q) - q E(q) = 0
    ,
    \label{eq:eqAIRY}
\end{equation}

\noindent where we have introduced the re-scaled spatial variable
\begin{equation}
    q \doteq x \left( \frac{\Omega^2}{c^2 L_n} \right)^{1/3}
    .
\end{equation}

Equation \eq{eq:eqAIRY} is known as Airy's equation, and contains a fold-type $A_2$ caustic at the cutoff location $q = 0$. Assuming that $E(q \to \infty) = 0$, the exact solution is given by the Airy function
\begin{equation}
    E_\textrm{ex}(q) = \airyA(q)
    ,
    \label{eq:exactAIRY}
\end{equation}

\noindent (where the overall constant is set to unity for simplicity) while the MGO solution \eq{eq:MGO} to \Eq{eq:eqAIRY} can be written in the underdense region $q \le 0$ as~\cite{Lopez20a}
\begin{align}
    E_\textrm{MGO}(q) =&
    \Upsilon\left(|q|^{1/2} \right) 
    \exp\left(- i \frac{2}{3} |q|^{3/2} \right) 
    \nonumber\\
    &+
    \Upsilon\left(-|q|^{1/2} \right) 
    \exp\left(i \frac{2}{3} |q|^{3/2} \right)
    .
    \label{eq:mgoAIRY}
\end{align}

\noindent The integral function $\Upsilon$ in \Eq{eq:mgoAIRY} has the form
\begin{equation}
    \Upsilon(p) 
    \doteq \frac{1}{2\pi}
    \int_{\cont{0}} \dd \epsilon \,
    \frac{
        \vartheta(p) 
        \exp\left[
        i f(\epsilon, p)
    \right]
    }{
        \left[ 
            \vartheta^4(p) - 8 \vartheta(p) p \epsilon 
        \right]^{1/4}
    }
    ,
    \label{eq:upsilonAIRY}
\end{equation}

\noindent where the phase function $f$ is given as
\begin{align}
    f(\epsilon,p) 
    &\doteq
    \frac{\vartheta^6(p) - \left[\vartheta^4(p) - 8 \vartheta(p) p \epsilon \right]^{3/2}}{96p^3}
    \nonumber\\
    &\hspace{4mm}- \frac{\vartheta^3(p)}{8p^2} \epsilon
    + \frac{\vartheta^2(p)}{4p} \epsilon^2
    ,
    \label{eq:fAIRY}
\end{align}

\noindent and we have defined $\vartheta(p) \doteq \sqrt{1 + 4 p^2}$. When \Eq{eq:upsilonAIRY} is evaluated using the stationary-phase approximation, the standard GO approximation for \Eq{eq:eqAIRY} is obtained:
\begin{equation}
    E_\textrm{GO}(q) =
    \pi^{-1/2} |q|^{-1/4} 
    \sin\left(\frac{2}{3} |q|^{3/2} + \frac{\pi}{4} \right)
    .
    \label{eq:goAIRY}
\end{equation}

\noindent Clearly, the GO solution diverges at the caustic $q = 0$. Conversely, if \Eq{eq:fAIRY} is expanded to cubic order in $\epsilon$, then \Eq{eq:upsilonAIRY} can be evaluated along the steepest-descent contour to yield the approximate MGO solution~\cite{Lopez20a}
\begin{align}
    E_\textrm{approx}(q) &=
    \sqrt{1 - 4 q} \, \airyA\left[- \varrho^2(q) \right]
    \cos[ \varpi(q)]
    \nonumber\\
    &\hspace{5mm}- \sqrt{1 - 4 q} \, \airyB\left[ - \varrho^2(q) \right]
    \sin[ \varpi(q) ]
     \, ,
     \label{eq:approxAIRY}
\end{align}

\noindent where we have defined
\begin{align}
    \varrho(q) &\doteq
    (1 - 4q) \sqrt{|q|}
    , \quad
    \varpi(q) \doteq 
    \frac{2}{3} \varrho^3(q) - \frac{2}{3}|q|^{3/2}
    .
\end{align}

\begin{figure}
    \centering
    \begin{overpic}[width=\linewidth,trim={4mm 5mm 4mm 4mm},clip]{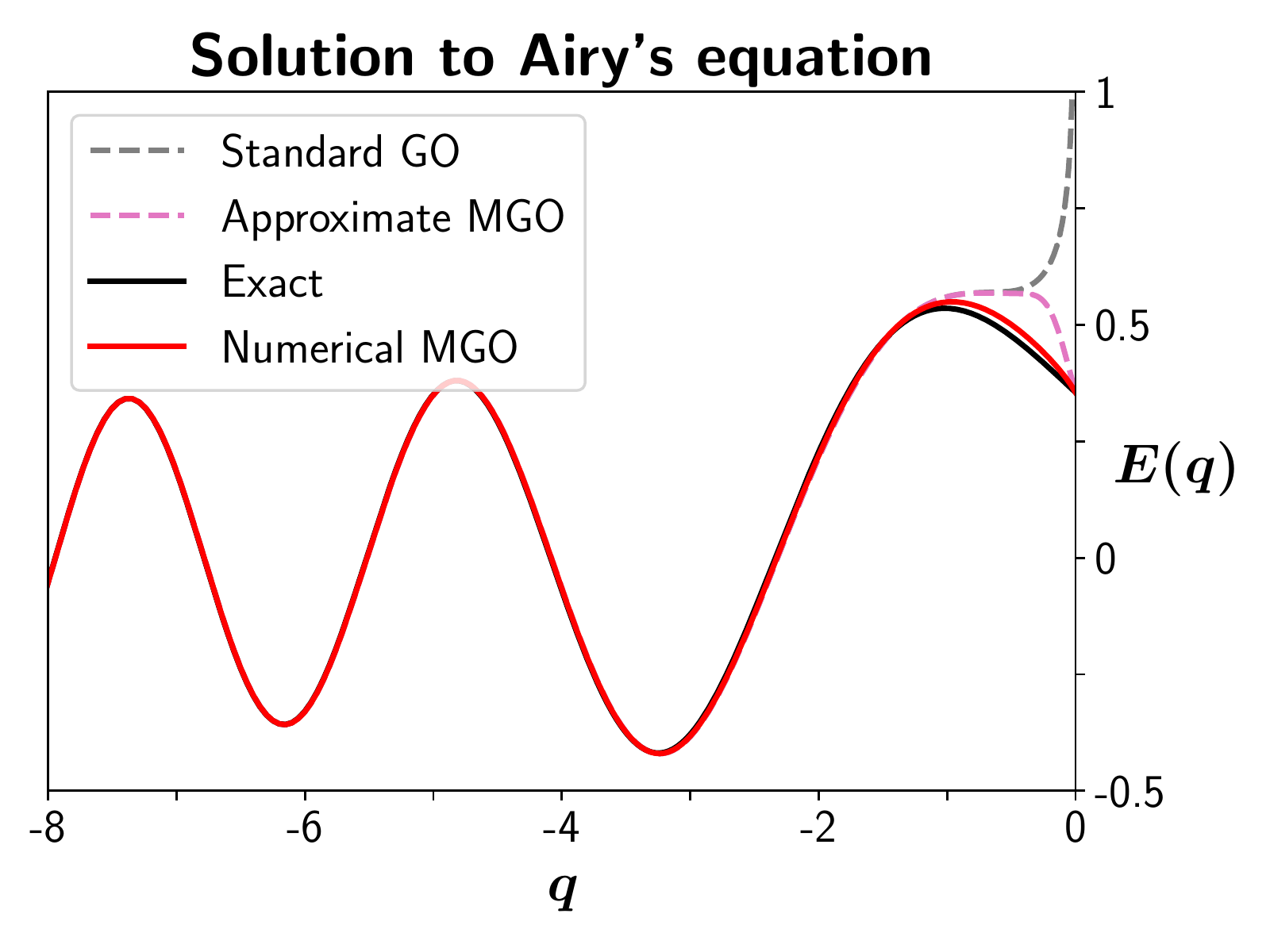}
        \put(5,12){\textbf{\large(a)}}
    \end{overpic}
    
    \vspace{2mm}
    \hspace{-3mm}\begin{overpic}[width=0.95\linewidth,trim={4mm 22mm 4mm 36mm},clip]{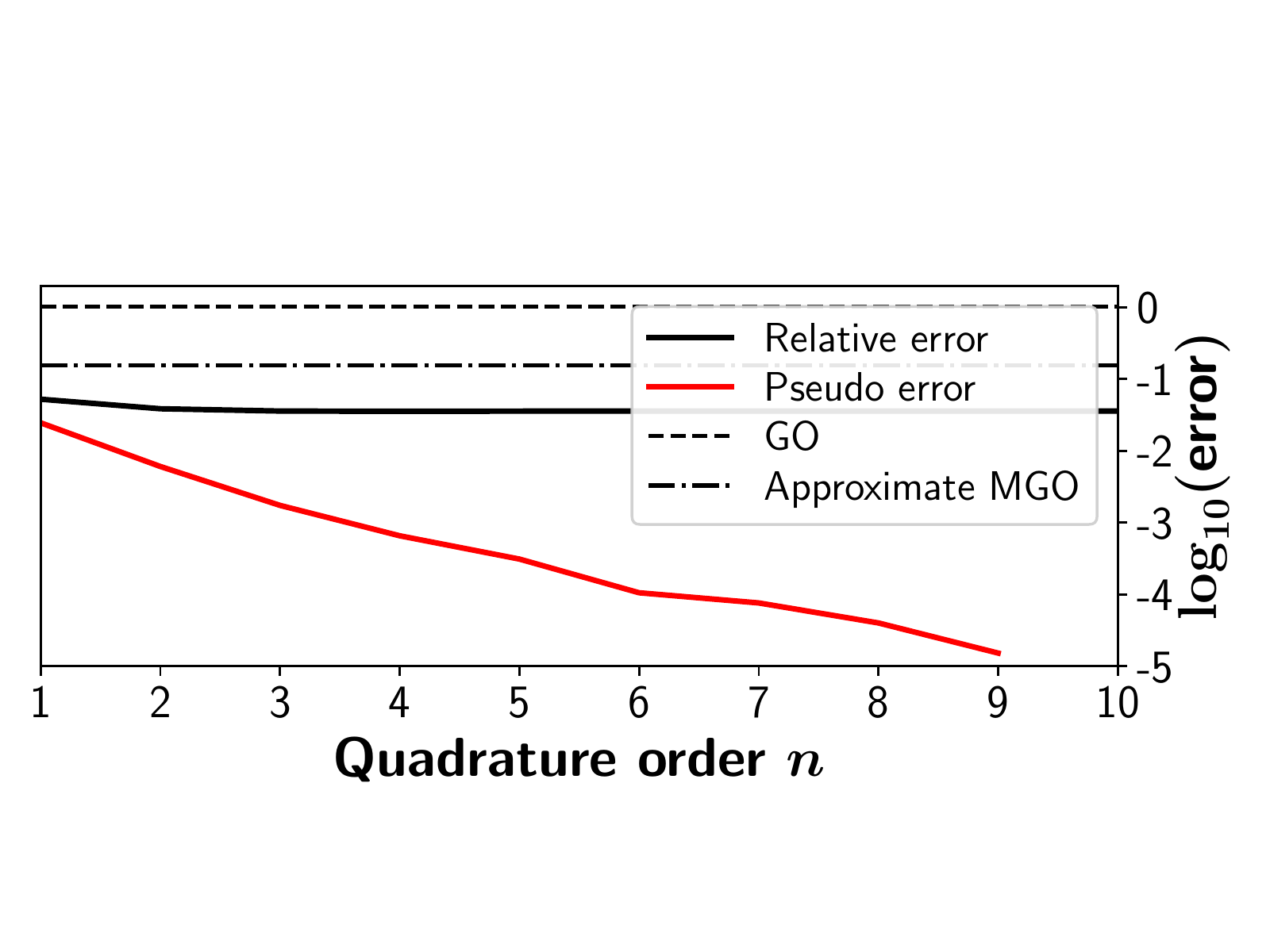}
        \put(5,12){\textbf{\large(b)}}
    \end{overpic}
    \caption{\textbf{(a)} Comparison of the numerical MGO solution (orange) with the analytically approximated MGO solution \eq{eq:approxAIRY}~\cite{Lopez20a} (dashed pink), the standard GO solution \eq{eq:goAIRY} (dashed gray), and the exact solution \eq{eq:exactAIRY} (black) for Airy's equation \eq{eq:eqAIRY}. The numerical MGO solution was obtained by applying the quadrature rule of \Eq{eq:MGOquad} with order $n = 10$ to \Eqs{eq:mgoAIRY} and \eq{eq:upsilonAIRY}. The numerical MGO solution displays remarkable agreement with the exact solution compared with the analytical approximations, even near the fold-type caustic at $q = 0$. \textbf{(b)} Error of the numerical MGO solution with respect to a scan over the quadrature order $n$. Note that the `pseudo error' is defined as the relative error with respect to the $n = 10$ solution used in \textbf{(a)}.}
    \label{fig:MGOairy}
\end{figure}

Here, we evaluate \Eq{eq:upsilonAIRY} numerically via \Eq{eq:MGOquad} over the range $q \in [-8, 0]$ using the angle memory feedback algorithm described in \Sec{sec:feedback}. Figure~\ref{fig:MGOcontour} shows the smooth evolution of steepest-descent curves obtained with the memory feedback algorithm, while \Fig{fig:MGOairy} compares the resultant numerical MGO solution with the exact solution \eq{eq:exactAIRY} and the two analytical approximations of  \Eqs{eq:goAIRY} and \eq{eq:approxAIRY}. As \Fig{fig:MGOairy} shows, both the numerical MGO solution and the analytically approximated MGO solution remain finite at the caustic $q = 0$, whereas the GO solution diverges. However, the analytical approximation overestimates the peak intensity width near the caustic. Conversely, the numerical MGO solution agrees remarkably well with the exact solution everywhere, even though a relatively low quadrature order of $n = 10$ was used. Moreover, although the relative error with respect to the exact solution does not decrease much after quadrature order $n = 2$, the `pseudo error' (defined as the relative error between the numerical MGO solution for a given $n$ compared with the reference solution $n = 10$) continues to decrease with increasing $n$. This suggests that the numerical MGO algorithm quickly converges to the residual intrinsic error of the MGO theory, at least for this specific example.


\section{Conclusions}

\label{sec:concl}

Metaplectic geometrical optics is a recently proposed formalism for modeling wave propagation in general linear media that avoids the usual singularities at caustics. MGO is therefore a promising alternative to the traditional GO approximation underlying ray-tracing codes. However, MGO yields solutions in the form of highly oscillatory integrals, which cannot be easily calculated using standard numerical methods. Here, we present a new algorithm for taking such integrals numerically that is based on the steepest-descent method combined with Gauss--Freud quadrature.

We first validate our algorithm on isolated saddlepoints of various degeneracy to demonstrate the expected $2n-1$ polynomial accuracy of an $n$-point Gaussian quadrature formula. We then use our algorithm to simulate an EM wave propagating into an unmagnetized plasma that has a fold-type caustic at the critical cutoff density. The numerical solution agrees remarkably well with the exact solution and significantly improves upon the analytically approximated MGO solution that was previously obtained in \Ref{Lopez20a}. This encouraging result provides strong evidence that MGO can be suitable for practical applications.

\section*{Acknowledgments}
The authors thank Laura Xin Zhang for invaluable coding advice. This work was supported by the U.S.~DOE through Contract No.~DE-AC02-09CH11466 and through funding for the Summer Undergraduate Laboratory Internship (SULI) program.


\appendix

\begin{table*}[t!]
    \centering
    \begin{tabular}{| c | c | c | c | c | c | c |}
        \multicolumn{1}{c}{Order} & \multicolumn{1}{c}{Nodes} & \multicolumn{1}{c}{Weights} & \multicolumn{1}{c}{} & \multicolumn{1}{c}{Order} & \multicolumn{1}{c}{Nodes} & \multicolumn{1}{c}{Weights} \\
        \hline
        \bf{n = 1} & 5.64189583547756 (1) & 8.86226925452758 (1) & & 
                   & 5.29786439318514 (2) & 1.34109188453360 (1) \\
        \cline{1-3}
                   & 3.00193931060839 (1) & 6.40529179684379 (1) & &
                   & 2.67398372167767 (1) & 2.68330754472640 (1) \\
        \bf{n = 2} & 1.25242104533372 (0) & 2.45697745768379 (1) & &
                   & 6.16302884182402 (1) & 2.75953397988422 (1) \\
        \cline{1-3}
                   & 1.90554149798192 (1) & 4.46029770466658 (1) & &
                   & 1.06424631211623 (0) & 1.57448282618790 (1) \\
        \bf{n = 3} & 8.48251867544577 (1) & 3.96468266998335 (1) & &
        \bf{n = 8} & 1.58885586227006 (0) & 4.48141099174625 (2) \\
                   & 1.79977657841573 (0) & 4.37288879877644 (2) & &
                   & 2.18392115309586 (0) & 5.36793575602526 (3) \\
        \cline{1-3}
                   & 1.33776446996068 (1) & 3.25302999756919 (1) & &
                   & 2.86313388370808 (0) & 2.02063649132407 (4) \\
                   & 6.24324690187190 (1) & 4.21107101852062 (1) & &
                   & 3.68600716272440 (0) & 1.19259692659532 (6) \\
        \cline{5-7}
        \bf{n = 4} & 1.34253782564499 (0) & 1.33442500357520 (1) & &
                   & 4.49390308011934 (2) & 1.14088970242118 (1) \\
                   & 2.26266447701036 (0) & 6.37432348625728 (3) & &
                   & 2.28605305560535 (1) & 2.35940791223685 (1) \\
        \cline{1-3}
                   & 1.00242151968216 (1) & 2.48406152028443 (1) & &
                   & 5.32195844331646 (1) & 2.66425473630253 (1) \\
                   & 4.82813966046201 (1) & 3.92331066652399 (1) & &
                   & 9.27280745338081 (1) & 1.83251679101663 (1) \\
        \bf{n = 5} & 1.06094982152572 (0) & 2.11418193076057 (1) & &
        \bf{n = 9} & 1.39292385519588 (0) & 7.13440493066916 (2) \\
                   & 1.77972941852026 (0) & 3.32466603513439 (2) & &
                   & 1.91884309919743 (0) & 1.39814184155604 (2) \\
                   & 2.66976035608766 (0) & 8.24853344515628 (4) & &
                   & 2.50624783400574 (0) & 1.16385272078519 (3) \\
        \cline{1-3}
                   & 7.86006594130979 (2) & 1.96849675488598 (1) & &
                   & 3.17269213348124 (0) & 3.05670214897831 (5) \\
                   & 3.86739410270631 (1) & 3.49154201525395 (1) & &
                   & 3.97889886978978 (0) & 1.23790511337496 (7) \\
        \cline{5-7}
                   & 8.66429471682044 (1) & 2.57259520584421 (1) & & 
                   & 3.87385243257289 (2) & 9.85520975191087 (2) \\
        \bf{n = 6} & 1.46569804966352 (0) & 7.60131375840058 (2) & & 
                   & 1.98233304013083 (1) & 2.08678066608185 (1) \\
                   & 2.17270779693900 (0) & 6.85191862513596 (3) & & 
                   & 4.65201111814767 (1) & 2.52051688403761 (1) \\
                   & 3.03682016932287 (0) & 9.84716452019267 (5) & & 
                   & 8.16861885592273 (1) & 1.98684340038387 (1) \\
        \cline{1-3}
                   & 6.37164846067008 (2) & 1.60609965149261 (1) & & 
                   & 1.23454132402818 (0) & 9.71984227600620 (2) \\
                   & 3.18192018888619 (1) & 3.06319808158099 (1) & & 
        \bf{n = 10}& 1.70679814968913 (0) & 2.70244164355446 (2) \\
                   & 7.24198989258373 (1) & 2.75527141784905 (1) & & 
                   & 2.22994008892494 (0) & 3.80464962249537 (3) \\
        \bf{n = 7} & 1.23803559921509 (0) & 1.20630193130784 (1) & & 
                   & 2.80910374689875 (0) & 2.28886243044656 (4) \\
                   & 1.83852822027095 (0) & 2.18922863438067 (2) & & 
                   & 3.46387241949586 (0) & 4.34534479844469 (6) \\
                   & 2.53148815132768 (0) & 1.23644672831056 (3) & & 
                   & 4.25536180636608 (0) & 1.24773714817825 (8) \\
                   & 3.37345643012458 (0) & 1.10841575911059 (5) & &
                   &              &              \\
        \hline
    \end{tabular}
    \caption{Gauss--Freud quadrature nodes and weights for quadrature orders up to $10$. The notation $a$ $(b)$ denotes $a \times 10^{-b}$.}
    \label{tab:GFnodes}
\end{table*}

\section{Gauss--Freud quadrature nodes and weights}
\label{sec:FreudQUAD}

The Freud polynomials are the unique family of polynomials that are orthogonal with respect to the inner product
\begin{equation}
    \langle h_1, h_2 \rangle = 
    \int_0^\infty \dd \kappa \, 
    h_1(\kappa) h_2(\kappa) \exp(-\kappa^2)
    .
\end{equation}

\noindent Since the Freud polynomials are uncommon, the corresponding quadrature nodes $\{ \kappa_j \}$ and weights $\{w_j\}$ are not typically provided in standard software. Moreover, the definitions of $\{ \kappa_j \}$ \eq{eq:GQnodes} and $\{ w_j \}$ \eq{eq:GQweights} are not practical when the functional forms of $\{ p_\ell(\kappa) \}$ are unknown.

In this case, it is better to use the Golub--Welsch algorithm~\cite{Golub69}, which relies on the following eigenvalue relationship that $\{ \kappa_j \}$ and $\{w_j\}$ can be shown to satisfy~\cite[pp.~141--144]{Gil07}:
\begin{equation}
    \Mat{J}_n \Vect{\nu}_j = \kappa_j \Vect{\nu}_j
    ,
    \quad
    j = 1, \ldots, n .
    \label{eq:golubNODES}
\end{equation}

\noindent Here, $\Mat{J}_n$ is the symmetric tridiagonal $n \times n$ Jacobi matrix corresponding to the first $n$ members of $\{ p_\ell(\kappa) \}$~\cite[p.~82]{Olver10a}, \ie
\begin{equation}
    \Mat{J}_n = 
    \begin{pmatrix}
        a_0        & \sqrt{b_1} &                & \\
        \sqrt{b_1} &        a_1 &         \ddots & \\
                   &     \ddots &         \ddots & \sqrt{b_{n-1}} \\
                   &            & \sqrt{b_{n-1}} & a_{n-1}
    \end{pmatrix}
    ,
\end{equation}

\noindent with $a_\ell$ and $b_\ell$ being the coefficients of the three-term recurrence relation that the \textit{monic} family $\{ \fourier{p}_\ell(\kappa) \}$ satisfy:
\begin{equation}
    \fourier{p}_{\ell + 1}(\kappa) 
    = (\kappa + a_\ell)\fourier{p}_{\ell}(\kappa)
    + b_\ell \fourier{p}_{\ell - 1}(\kappa) 
    , \quad
    \ell = 0, 1, \ldots
\end{equation}

\noindent subject to the initial conditions
\begin{equation}
    \fourier{p}_{-1}(\kappa) = 0 
    , \quad
    \fourier{p}_0(\kappa) = 1 .
\end{equation}

\noindent There are established algorithms to obtain these coefficients~\cite{Press07gauss,Gautschi16}. The weights are then obtained from the first eigenvector $\Vect{\nu}_1$, which can be normalized such that $\{w_j\}$ are given by its vector components as
\begin{equation}
    \Vect{\nu}_1 = \frac{1}{\sqrt{\langle 1, 1 \rangle}}
    \begin{pmatrix}
        \sqrt{w_1}
        & 
        \ldots
        &
        \sqrt{w_n }
    \end{pmatrix}^\intercal
    , \quad
    \Vect{\nu}_1^\intercal \Vect{\nu}_1 = 1
    .
    \label{eq:golubWEIGHTS}
\end{equation}

The resulting list of $\{\kappa_j\}$ and weights $\{w_j\}$ for quadrature orders $n \le 10$ is provided in Table~\ref{tab:GFnodes}, adapted from a similar table for $2 \le n \le 20$ presented in \Ref{Steen69}. These values can also be calculated with high precision for arbitrary values of $n$ using the code of \Ref{Gautschi20}; see \Ref{Gautschi21} for more details.


\bibliography{bib.bib}
\bibliographystyle{apsrev4-1}

\end{document}